\documentclass[twocolumn,aps,superscriptaddress,nolongbibliography,pra,10pt]{revtex4-2}
\usepackage{graphicx}
\usepackage{amsmath}
\usepackage{amssymb}
\usepackage{bm}
\usepackage{hyperref}
\usepackage[dvipsnames]{xcolor}
\newcommand{\notleftright}{\mathrel{\ooalign{$\longleftrightarrow$\cr\hidewidth$/$\hidewidth}}}

\usepackage[normalem]{ulem}

\begin{document}

\title{Ro-vibrational van der Waals interaction between ultracold polar molecules}
\author{Kang Feng, Hanwei Yang, Hubert J. J\'{o}\'{z}wiak, Tijs Karman}
\affiliation{Institute for Molecules and Materials, Radboud University, Nijmegen, The Netherlands}

\begin{abstract}
We describe the ro-vibrational van der Waals interaction between ultracold polar molecules.
This interaction is strong, leading to fast elastic collisions and orders of magnitude suppression of collisional loss.
This enables evaporative cooling of Fermi mixtures of molecules in different ro-vibrational states,
without active shielding by applying external fields.
The scheme is compatible with microwave shielding,
where it enables controlled state dependent interactions,
opening up new opportunities for quantum simulation and impurity physics.
The interaction can also be used to stabilize fermionic molecules in optical lattices,
to control interactions in synthetic dimensions,
for enhanced tweezer loading, and direct infrared shielding.
\end{abstract}

\maketitle

\section{Introduction}

%general promise and land mark papers
Ultracold polar molecules are long considered powerful platforms for few~\cite{karman:24} and many-body physics~\cite{cooper:09,gorshkov2011}, quantum simulation~\cite{micheli2006toolbox,cornish:24} and computation~\cite{demille2002quantum,ruttley2025long,picard2025entanglement,holland2023demand,bao2023dipolar}, and precision measurement~\cite{demille2024quantum,Safronova2018}.
While quantum control of molecules is challenging compared to atoms,
recent progress has been rapid~\cite{langen2024quantum}.
Molecules have anisotropic, tunable dipole–dipole interactions that are long ranged and extend over multiple sites in an optical lattice~\cite{wall2010hyperfine,trefzger2011ultracold}.
These interactions enable the natural realization of extended Hubbard models~\cite{capogrosso2010quantum}, quantum magnetism with controllable spin–spin interactions~\cite{Barnett2006, micheli2006toolbox, gorshkov2011,carroll2025observation}, and supersolids~\cite{zhang2025supersolid}, quantum droplets~\cite{langen2025dipolar}, and other novel quantum phases~\cite{ciardi2025self}.
In addition, molecules provide a large internal Hilbert space through their rotational, vibrational, and hyperfine degrees of freedom.
This allows quantum simulators to encode pseudo-spin and multi-orbital physics,
SU($N$) symmetric interactions~\cite{mukherjee2025n},
and synthetic dimensions~\cite{sundar2018synthetic,feng2022quantum}

%shielding has recently yielded quantum gases with controllable interactions
Shielding has recently yielded quantum gases of molecules with tunable interactions,
both Fermi degenerate gases~\cite{demarco:19,valtolina:20,matsuda:20,schindewolf:22} and Bose-Einstein condensates~\cite{bigagli:24,shi2025boseeinsteincondensateultracoldsodiumrubidium} of polar molecules.
Shielding is necessary for collisional stability~\cite{yuan:26} and efficient evaporative cooling,
but simultaneously offers control over molecular interactions.
Double microwave shielding offers control of the sign and strength of dipolar interactions,
and simultaneously enables tuning of the scattering length relative to the dipolar one~\cite{karman:25}.
Control of the microwave ellipticity offers further control of the anisotropy of the dipolar interactions~\cite{chen:23,zhang:26}.
Quantum gases of polar molecules with tunable interactions offer a powerful platform for quantum many-body physics and simulation~\cite{schindewolf2025few}.

In this context, additional schemes for controlling collisions and interactions can be highly valuable as a resource for interaction tuning.
We have previously developed the rotational~\cite{walraven:24a} and hyperfine~\cite{walraven:25} van der Waals interactions.
These interactions can be repulsive and hence suppress adverse collisions,
and they can be controlled through the state dependence.
Unlike most collisional shielding schemes, these interactions can be effective even in the absence of external fields.

%Rotational van der Waals~\cite{walraven:24a}
These state-dependent van der Waals interactions are qualitatively understood as follows.
The dipole-dipole interaction is the dominant long-range interaction between polar molecules,
scaling with intermolecular distance as $R^{-3}$.
However, without external fields any eigenstate of the molecule can be assigned definite parity and has no dipole expectation value.
The dipolar interaction is still the dominant interaction,
but its effect stems from the transition dipole moment between molecular eigenstates of opposite parity.
These virtual transitions lead to a second-order interaction which varies as $R^{-6}$ and is typically called a van der Waals (vdW) interaction.
Selection rules determine which virtual excitations contribute to this interaction, i.e. which have non-zero transition dipole.
The second-order interaction is typically dominated by the allowed virtual transition with the smallest energy denominator,
and the sign of the energy denominator determines the sign of the interaction;
dominant virtual excitations (de-excitations) lead to attractive (repulsive) interactions.
Thus, the internal states of the molecules control the strength and sign of the vdW interaction \cite{walraven:24a}.

For rotational states of diatomic molecules, the dipole operator has the selection rule $\Delta j =\pm1$.
Two molecules will experience resonant dipole-dipole interactions when placed in rotational states $j,j+1$, which is coupled by dipole-dipole to $j+1,j$.
When molecules are placed in any other choice of rotational states $j,j'$,
dipole-dipole couples this to pairs of states $j\pm1,j'\pm1$.
The energy denominators are $j,j'$ dependent, and are on the order of the rotational constant.
This results in a second-order interaction that we called the rotational van der Waals~\cite{walraven:24a} to indicate that the dominant virtual excitations are purely rotational transitions,
in contrast with the conventional \emph{electronic} vdW interaction.
The rotational vdW interaction is state dependent through $j$ and $j'$,
and repulsive whenever $\Delta j \ge 2$,
which results directly from the $j(j+1)$ energy spectrum of the rigid rotor that gets sparser as energy increases.
We note that this is contrary to electronic energy levels which typically become denser at higher energies and for which the vdW interaction is typically attractive.

%Hyperfine van der Waals\cite{walraven:25}
The hyperfine van der Waals interaction~\cite{walraven:25} occurs for the special case where $j'=j+1$.
In this case, one should nominally observe resonant dipolar interactions.
However, if one considers molecules in pairs of hyperfine states labeled by rotational and total angular momentum $j,f$,
one can find pairs of states $(j,f)+(j',f')$ such that even though $j \rightarrow j'=j+1$ transitions are dipole allowed, $f\notleftright f'$ transitions are not,
e.g. for $\Delta f > 1$ or $f=0 \notleftright f'=0$.
In this case, the dipolar interaction couples this pair of states of the molecules to other hyperfine states within the same $j,j'$ manifold.
The energy denominator for such virtual excitations is not on the order of the rotational constant as for the rotational vdW interaction,
but on the other of the hyperfine splittings.
Hence, the energy denominator is orders of magnitude smaller and the corresponding van der Waals interaction orders of magnitude stronger.
This interaction is not effective for closed-shell molecules where the hyperfine splittings are too small,
but it is very effective for $^2\Sigma$ molecules which includes most laser-coolable molecules~\cite{walraven:25}.

\begin{figure}
    \centering
    \includegraphics[width=1.0\linewidth]{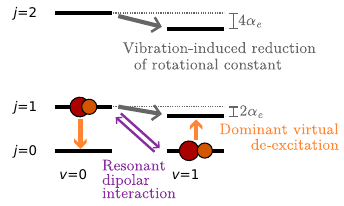}
    \caption{
    {\bf Molecular energy levels relevant for the ro-vibrational vdW interaction.}
    The rotation-vibration coupling leads to a reduction of the rotational constant in vibrationally excited states, as indicated in gray.
    This lifts the resonances of the rotational de-excitation in $v=0,j=1$, and rotational excitation in $v=1,j=0$,
    indicated by the orange arrows.
    This virtual process dominates the second-order dipolar interaction that gives rise to the ro-vibrational vdW interaction.
    The purple arrows indicate that a direct resonant dipole-dipole interaction also exists,
    but this involves the vibrational transition dipole moment that is at least an order of magnitude weaker than the rotational one for the molecules that we consider.}
    \label{fig:cartoon}
\end{figure}

% ro-vibrational vdW
In this paper, we detail a novel ro-vibrational van der Waals interaction, which is illustrated in Fig.~\ref{fig:cartoon}.
We consider one molecule in the vibrational ground state $v=0$ and rotational state $j=1$,
and one molecule in the first vibrationally excited state $v'=1$ and rotational state $j'=0$.
If the rotational constant were the same in both vibrational states,
this pair $(v,j)+(v',j') = (0,1)+(1,0)$ would be degenerate with $(0,0)+(1,1)$.
These are related by a dipole-allowed purely rotational transition for both molecules,
and hence coupled by the dipolar interaction.
However, due to the anharmonicity of the molecular vibration, the molecular bond is slightly stretched in the vibrationally excited state,
which leads to a slight reduction of the rotational constant that makes the dipolar interaction non-resonant, see Sec.~\ref{sec:dunham}.
The typical energy defect is in the order of 10~MHz,
resulting in a vdW interaction that is orders of magnitude stronger than the rotational counterpart.

%vibrations
Vibration --unlike rotation-- is not currently exploited as a resource of ultracold molecules~\cite{cornish:24},
with very few exceptions including attempts to control chemical reactivity~\cite{ye2018collisions} and using $\ell$-type doubling for precision measurement~\cite{kozyryev2017precision,hutzler2020polyatomic,anderegg2023quantum} and shielding~\cite{augustovivcova2019ultracold,vilas2026quantum}.
Nevertheless, ultracold molecules could readily be prepared in vibrationally excited states.
Typical ultracold molecules are bialkalis produced by associating ultracold atoms using stimulated Raman adiabatic passage (STIRAP) \cite{ni2008high,takekoshi2014ultracold,molony2014creation,park2015ultracold,guo2016creation,Stevenson2023}.
STIRAP can produce molecules in the vibrational ground state or directly in a vibrationally excited state~\cite{ye2018collisions}.
Alternatively, the STIRAP laser and a single additional laser can drive a Raman transition into the vibrationally excited state.
For laser-coolable molecules, such as CaF~\cite{zhelyazkova2014laser}, SrF~\cite{shuman2010laser}, BaF~\cite{rockenhauser2024laser}, and YO~\cite{collopy20183d}, vibrationally excited states can be prepared by turning off a vibrational repump laser or a particular hyperfine component therein.
Alternatively, the main cooling laser and a vibrational repump laser can be used to drive a Raman transition.

While the ro-vibrational vdW interaction is similar in origin to the known rotational and hyperfine vdW interactions, we believe that this can be a powerful, versatile resource for controlling interactions between ultracold molecules. To demonstrate this, the manuscript is structured as follows. We begin in Sec.~\ref{sec:dunham} by detailing the origin of this rotation-vibration energy defect and compile these coupling constants for relevant molecules. In Sec.~\ref{sec:rvdW}, we qualitatively characterize the ro-vibrational vdW interaction, comparing its energy and length scales with its electronic, rotational, and hyperfine counterparts. In Sec.~\ref{sec:CC}, we extend coupled-channel calculations to include molecular vibration, enabling the quantitative study of collision dynamics.
The results, discussed in Sec.~\ref{sec:results}, confirm that
the ro-vibrational vdW interaction is stronger, leading to faster elastic collisions and better suppression of inelastic losses.
%, when compared to its rotational or hyperfine counterpart.
In Sec.~\ref{sec:universality}, we demonstrate the universality of the ro-vibrational vdW interaction.
In Sec.~\ref{sec:evap}, we consider direct evaporative cooling of fermionic mixtures without active shielding,
enabled by the favorable collisional properties.
We conclude in Sec.~\ref{sec:discussion} by discussing further applications enabled by this work,
including direct collisional shielding,
enhanced deterministic loading of molecular arrays,
and compatibility with microwave shielding, which opens up new opportunities for quantum magnetism and impurity physics.

\section{Rotation-vibration coupling \label{sec:dunham}}

The rotation-vibration energy levels of a diatomic molecule are described by the Dunham expansion~\cite{dunham:32}, a power series in $j(j+1)$ and $v+1/2$,
\begin{align}
    E(v,j) &= B_e j\left(j+1\right) - D_e \left[j\left(j+1\right)\right]^2 + \ldots \nonumber \\
    +& \omega_e \left(v+\frac{1}{2}\right) - \omega_ex_e\left(v+\frac{1}{2}\right)^2 + \ldots \nonumber \\
    -& \alpha_e j\left(j+1\right) \left(v+\frac{1}{2}\right) + \ldots.
    \label{eq:dunham}
\end{align}
The first term describes the energy levels of a rigid rotor, where $B_e$ is the rotational constant.
The second term describes rotational distortion, lessening of the rotational constant upon rotational excitation, where $D_e$ is the rotational distortion constant.
The third term describes harmonic vibrational energy levels, and $\omega_e$ is the harmonic oscillator frequency.
The fourth term describes the anharmonicity, quantified by $\omega_ex_e$.
The fifth term describes the rotation-vibration coupling, and $\alpha_e$ is the rotation-vibration coupling constant.
Essentially, this describes reduction of the rotational constant upon vibrational excitation,
which effectively increases the mean bond length and hence the molecule's moment of inertia.
The sign of each term in the Dunham expansion is chosen such that the molecular constants are typically positive, but as we will discuss below this is somewhat subtle for the case of the rotation-vibration coupling constant $\alpha_e.$

Pekeris derived an expression for the rotation-vibration coupling constant $\alpha$ for the Morse potential\cite{pekeris:34}
\begin{subequations}\label{eq:pekeris_all}
\begin{align}
\alpha_e &= 6 B_e \left( \sqrt{\frac{B_ex_e}{\omega_e}} - \frac{B_e}{\omega_e} \right) \label{eq:pekeris1} \\
&= 3 \sqrt{\omega_e x_e D_e} - 6 \frac{B_e^2}{\omega_e} \label{eq:pekeris2} \\
&= 3 B_e \sqrt{\frac{B_e}{V}} - 6 \frac{B_e^2}{\omega_e}. \label{eq:pekeris3}
\end{align}
\end{subequations}
The first equation here is Pekeris' relation\cite{pekeris:34},
the second is related to it by the Kratzer relation $D_e = 4B_e^3/\omega_e^2$, which holds for many reasonable potentials,
and the third is related by $V = \omega_e/4x_e$, a relation for the potential depth that holds for the Morse potential.
In all three cases, the second term is the result for a purely harmonic potential.
In the harmonic case, the vibrational wavefunction becomes wider upon vibrational excitation,
but remains centered around $r_e$,
and the rotational constant related to the expectation value of $1/r^2$ increases.
For real anharmonic molecular potentials, the vibrational wavefunction not only becomes wider upon vibrational excitation,
but it also becomes asymmetric;
The wavefunction amplitude grows more on the large $r$ side and the mean interatomic distance grows,
resulting in a decrease of the rotational constant upon vibrational excitation.
Hence, the sign of $\alpha$ reverses from the harmonic to the anharmonic case,
such that the first term of the Pekeris relation typically dominates, but only by a factor a few.
This already indicates that $\alpha_e$ is quite sensitive to the precise form of the potential anharmonicity.
For example, for the Kratzer potential $2V[(r_e/r)^2-r_e/r]$~\cite{kratzer:20}, $\alpha_e=2B_ex_e$ which is different from the Pekeris relation by a factor $3\sqrt{B_e/\omega_e x_e} - 3B_e/\omega_ex_e$ which is around 0.67 for many molecules considered by Pekeris~\cite{pekeris:34}.
Interestingly, this expression was more recently rederived as an approximate result for the Morse potential\cite{burkhardt:07}, although as we just discussed it was already known to be inaccurate compared to the exact result by Pekeris\cite{pekeris:34}.

We tabulate the rotation-vibration coupling constant $\alpha_e$ for various molecules in Table~\ref{tab:constants}.
For bi-alkali molecules we determine these by numerically computing ro-vibrational wavefunctions using spectroscopically accurate potential energy curves~\cite{docenko:04,docenko:11,pashov:05,pashov:07,steinke:12,ferber:09,ivanova:11,tiemann:09}, and fitting a Dunham expansion.
We also include some laser-coolable molecules for which $\alpha_e$ is known directly from spectroscopy.
It can be seen that the values obtained for $\alpha_e$ agree reasonably but only qualitatively with the Pekeris relations Eqs.~\eqref{eq:pekeris_all}.
Values resulting from Eqs.~\eqref{eq:pekeris1} and \eqref{eq:pekeris2} agree well,
because these are related by the Kratzer relation that is widely considered to be reasonable for many realistic potential models.
Values obtained from Eq.~\eqref{eq:pekeris3} appear to be less accurate but qualitatively meaningful (except for the case of LiK),
while this relation has the advantage that it does not require knowledge of $\omega_ex_e$.
The difference between these estimates seems to be indicative of the difference with $\alpha_e$ from experiment, where avaiable.
We emphasize that Eq.~\eqref{eq:pekeris3} differs from Eq.~\eqref{eq:pekeris1} only by $V = \omega_e/4x_e$, which holds exactly for the Morse potential for which the Pekeris relation~\eqref{eq:pekeris1} was derived in the first place.
Hence, the relatively better performance of one relation over the others is an empirical observation that does not need to hold for all molecules.
There is also substantial interest in ultrapolar molecules consisting of alkali metal and coinage metal atoms, for which spectroscopic parameters are computed in Ref.~\cite{Smialkowski}.
From this data available, we estimated the rotational distortion by the Kratzer relation $D_e = 4B_e^3/\omega_e^2$, the anharmonicity as $\omega_ex_e = \omega_e^2/4V$, and the rotation-vibration coupling constant by Eq.~\eqref{eq:pekeris3}.
From Eqs.~\eqref{eq:pekeris_all}, a strong dependence on molecular parameters is apparent,
which explains the values for $\alpha_e$ that span orders of magnitude, and are typically between 1~MHz to 100~MHz for typical \emph{assembled} ultracold molecules.

\begin{table*}
    \centering
    \caption{{\bf Molecular constants}. For the bialkalis all molecular parameters are obtained by fitting a Dunham expansion to the ro-vibrational states calculated on the spectroscopically accurate potential energy curves, as indicated. Dipole moments and derivatives are taken from Refs.~\cite{aymar2005calculation} and~\cite{Ladjimi}, as indicated. For the coinage-metal molecules, parameters are taken from Ref.~\cite{Smialkowski}, while the rotational distortion was estimated from these parameters by the Kratzer relation $D_e = 4B_e^3/\omega_e^2$, the anharmonicity estimated as $\omega_ex_e = \omega_e^2/4V$, and $\alpha_e$ estimated using the third Pekeris relation, Eq.~\eqref{eq:pekeris3}.
    For the laser-coolable molecules, spectroscopic parameters are taken directly from literature as indicated.
    }
    \begin{tabular}{lcccccccccc}
        \hline\hline
        Molecule & $d_e$ (Debye) & $\frac{\partial d}{\partial r}$ (Debye/$a_0$) & $B_e$ (GHz) & $D_e$ (kHz) & $\omega_e$ (THz) & $\omega_ex_e$ (GHz) & $\alpha_e$ (MHz) Expt. & Eq.~\eqref{eq:pekeris1} & Eq.~\eqref{eq:pekeris2} & Eq.~\eqref{eq:pekeris3}  \\
        \hline
        $^{14}$NH            &  1.53~\cite{koput2015ab} & 0.37~\cite{koput2015ab}  & 500.63 & 51,\ 255 & 98.4 & 2\,347 &  19\, 460 & 17\,810 & 17\,620 &~\cite{radford1975imine,wayne1976laser} \\
        $^{24}$Mg$^{19}$F    &  3.6~\cite{torring1984dipole}  & 2.6~\cite{pasteka} & 15.6 & 32.4 & 21.3 & 148.1 & 140.9 & 142.0 & 139.6 &~\cite{Huber_1979}\\
        $^{40}$Ca$^{19}$F    &  3.1~\cite{CHILDS1986215}  &  3.1~\cite{hao} & 10.3 & 13.5 & 17.4 & 82.1 & 73.6 & 65.3 & 64.5 &~\cite{anderson:94,Huber_1979} \\
        $^{88}$Sr$^{19}$F    &  3.5~\cite{ernst1985electric}  &  3.1~\cite{hao}  & 7.51 & 7.46 & 15.1 & 66.0 & 46.3 & 44.12 & 44.09 &~\cite{Huber_1979}\\
        $^{138}$Ba$^{19}$F   &  3.4~\cite{torring1984dipole}  &  3.5~\cite{hao}  & 6.47 & 5.25 & 14.1 & 53.7 & 36.0 & 33.6 & 32.5 &~\cite{Huber_1979}\\
        $^{89}$Y$^{16}$O     &  4.5~\cite{suenram}  &    & 11.6 & 9.59 & 25.8 & 86.9 & 54.0 & 54.5 & 55.2 &~\cite{Huber_1979}\\
        LiNa                 &  0.47~\cite{Ladjimi}  &  0.049~\cite{aymar2005calculation}  & 11.3 & 98.2 & 7.69 & 49.5 & 96.2 & 108 & 110 & 147~\cite{steinke:12} \\
        LiK                  &  3.4~\cite{Ladjimi}  &  0.19~\cite{aymar2005calculation}  & 7.72 & 47.3 & 6.36 & 38.3 & 58.3 & 69.1 & 71.6 & -56.1~\cite{tiemann:09} \\
        LiRb                 &  4.0~\cite{Ladjimi}  &  0.22~\cite{aymar2005calculation}  & 6.48 & 32.3 & 5.87 & 43.0 & 47.8 & 67.6 & 68.8 & 74.5~\cite{ivanova:11} \\
        LiCs                 &  5.3~\cite{Ladjimi}  &  0.33~\cite{aymar2005calculation}  & 5.63 & 23.6 & 5.53 & 29.9 & 38.0 & 44.9 & 45.3 & 61.1~\cite{staanum:07} \\
        NaK                  &  2.7~\cite{Ladjimi} &   0.10~\cite{aymar2005calculation} & 2.85 & 6.80 & 3.72 & 14.9 & 13.7 & 16.9 & 17.1 & 24.4~\cite{hartmann:19} \\
        NaRb                 &  3.3~\cite{Ladjimi} &   0.12~\cite{aymar2005calculation} & 2.09 & 3.66 & 3.23 & 13.3 & 9.0  & 12.4 & 12.8 & 15.3~\cite{pashov:05} \\
        NaCs                 &  4.7~\cite{Ladjimi} &   0.20~\cite{aymar2005calculation} & 1.74 & 2.42 & 2.96 & 9.93 & 7.09 & 8.51 & 8.58 & 11.7~\cite{docenko:04} \\
        KRb                  &  0.63~\cite{Ladjimi} &  0.0089~\cite{aymar2005calculation}  & 1.14 & 1.16 & 2.27 & 6.91 & 3.67 & 5.01 & 5.07  & 6.79~\cite{pashov:07} \\
        KCs                  &  1.9~\cite{Ladjimi} &   0.059~\cite{aymar2005calculation} & 0.91 & 0.73   & 2.05 & 5.83 & 2.73 & 3.73 & 3.76 & 5.06~\cite{ferber:09} \\
        RbCs                 &  1.2~\cite{Ladjimi} &   0.044~\cite{aymar2005calculation} & 0.49 & 0.15   & 1.78 & 4.68 & 1.26 & 1.69 & 1.71 & 1.86~\cite{docenko:11} \\
        LiAg                 & 5.21  & 1.0~\cite{Smialkowski} & 13.8 & 76.4 & 11.7 & 73.0 & & & & 127  ~\cite{Smialkowski} \\
        NaAg                 & 6.0   & 1.2~\cite{Smialkowski}   & 3.83 & 5.51 & 6.37 & 26.0 & & & & 22.1 ~\cite{Smialkowski} \\
        KAg                  & 8.5   & 1.4~\cite{Smialkowski} & 2.00 & 1.64 & 4.12 & 12.3 & & & & 8.1  ~\cite{Smialkowski} \\
        RbAg                 &  9.0  & 1.6~\cite{Smialkowski}   & 1.11 & 0.500 & 3.29 & 6.91 & & & & 3.35 ~\cite{Smialkowski} \\
        CsAg                 &  9.8  & 1.9~\cite{Smialkowski}   & 0.806 & 0.274 & 2.77 & 4.71 & & & & 2.00 ~\cite{Smialkowski} \\
        FrAg                 &  9.2  & 1.8~\cite{Smialkowski}   & 0.645 & 0.168 & 2.52 & 4.18 & & & & 1.53 ~\cite{Smialkowski} \\
        \hline\hline
    \end{tabular}
    \label{tab:constants}
\end{table*}

\section{Ro-vibrational van der Waals interaction \label{sec:rvdW}}

%vdW interaction
We consider dipolar interactions between molecules in different ro-vibrational states $v,j$ and $v',j'$.
First-order dipolar interactions occur only if the transition $v,j$ to $v',j'$ is dipole allowed,
so that the pair of states $(v,j)+(v',j')$ is coupled by the dipolar interaction to $(v',j')+(v,j)$.
Otherwise, the dominant interaction arises in second order, resulting in a van der Waals interaction
\begin{equation}
    \sum_{n'\neq n}\frac{\left|\langle n'|\hat{V}_\mathrm{dd}(R)|n\rangle\right|^2}{E_n-E_{n'}}=-\frac{C_6}{R^6},
    \label{eq:secondorder}
\end{equation}
where $n=(v,j)+(v',j')$ is a short-hand notation for a pair of states,
with energy $E_n$,
$\hat{V}_\mathrm{dd}(R)$ is the dipole-dipole interaction,
and $C_6$ the resulting van der Waals coefficient.
Note that a strong interaction is obtained for molecules in states where there is coupling to a virtual state $n'$ that is very close in energy, minimizing the energy denominator in Eq.~\eqref{eq:secondorder}.
The sign of the interaction is determined by this denominator:
dominant virtual states that are higher (lower) in energy result in attractive (repulsive) interactions.

% selection rules
Pairs of virtual excitations contribute to Eq.~\eqref{eq:secondorder} only if the transition is dipole-allowed for both molecules.
For pure rotational transitions this leads to the selection rule $\Delta j = \pm1$, $\Delta v=0$,
and the transition dipole moment is on the order of the permanent dipole moment of the molecule around the equilibrium bond length, $d_e$.
The selection rule for vibrational transitions is $\Delta j = \pm1$ and $\Delta v = \pm 1$,
and here the transition dipole moment is on the order of the derivative of the dipole with respect to bond length, $\partial d/\partial r$, around equilibrium.

% ro-vibrational vdW
In this paper, we study the special case where two molecules are prepared in $(v,j)+(v',j')=(0,1)+(1,0)$, as illustrated in Fig.~\ref{fig:cartoon}.
This pair of molecular states is coupled by the dipole-dipole interaction to $(0,0)+(1,1)$,
which is lower in energy by $2\alpha_e$.
This leads to a second-order interaction $\pm C_6 R^{-6}$ which is repulsive (attractive) in the upper (lower) threshold,
with $C_6 = d_e^4/9\alpha_e$ the isotropic vdW coefficient that determines the interaction in the $s$-wave channel.
This interaction can be very strong; $C_6 = 7 \times 10^7$ a.u. for NaK and $10^{10}$ a.u. for KAg,
which leads to corresponding length scales $R_6 = (mC_6/\hbar^2)^{1/4} = 1700$ and $7400$~$a_0$, respectively.
Although the molecular vibration is an essential element for this interaction mechanism,
we emphasize that this should not be considered a \textit{vibrational} vdW interaction, as the virtual transitions are purely rotational.
We therefore refer to this as the \textit{ro-vibrational} vdW interaction,
emphasizing that the rotation-vibration coupling is responsible for the energy denominator associated with this virtual excitation.

%typical numbers
Typical electronic excitation energies are in the eV range,
such that typical $C_6$ coefficients that characterize the strength of the van der Waals interaction $-C_6R^{-6}$ are in the order 100--$10\,000$ atomic units,
strongly depending on the size of the constituent atoms.
The strength of this van der Waals interaction can also be conveniently characterized by a length scale $R_6 = (mC_6/\hbar^2)^{1/4}$ which is in the order of a hundred atomic units.
Note that this length scale scales relatively weakly with $C_6$.
Typical rotational constants are in the order of several GHz,
leading to typical $C_6$ coefficients in the order of hundreds of thousands of atomic units,
and typical van der Waals lengths $R_6$ in the order of several hundred bohr radii.
For open-shell molecules for which the hyperfine van der Waals interaction is effective,
the fine and hyperfine splittings are on the order of tens to hundreds of MHz,
leading to $C_6$ coefficients in the millions of atomic units, 
and $R_6$ in the order of a thousand bohr radii.
Compared to these, the ro-vibrational vdW interaction discovered here is orders of magnitude stronger,
characterized by $C_6$ coefficients in the order of tens of millions to billions of atomic units,
and $R_6$ length scales of thousands of bohr radii.

% v>1
The mechanism occurs more generally for pairs of molecules in states $(v,j)+(v',j+1)$ that are connected by purely rotational transition dipoles to $(v,j+1)+(v',j)$,
which are degenerate except for the change of rotational constant in the different vibrational states.
The energy defect is $2\alpha_e (j+1)\Delta v$ and leads to repulsive (attractive) vdW interactions in the $(v,j)+(v',j+1)$ channel if $v>v'$ ($v<v'$).

For the molecules considered here, the transition dipole moment for vibrational transitions is at least an order of magnitude smaller than for rotational transitions.
This has two important consequences.
First, the vibrational transition dipole moment gives rise to first-order dipole-dipole interactions in the $(v,j)+(v',j') = (0,1)+(1,0)$ threshold, see Fig.~\ref{fig:cartoon}.
We have so far ignored this to simplify the discussion.
In reality, this dipole-dipole interaction arising from the vibrational transition dipole moment is so weak that it can be neglected, see Sec.~\ref{sec:universality}.
Second, the weak vibrational transition dipole moment also means that there is no effective \emph{vibrational} vdW interaction.
This seems obvious because the transition dipole moment is weaker than for the competing rotational vdW interaction,
and the energy denominator appears orders of magnitude larger, if it is on the order of $\omega_e$.
As described in the Appendix, however, the situation is slightly more subtle because the energy denominator for virtual processes of the type $v,v\rightarrow v-1,v+1$ is determined by the anharmonicity $\omega_ex_e$,
which is on the order of GHz, just like the rotational constant,
so that it is possible to cancel the energy denominator to a large extent.
Due to the small vibrational transition dipole moment, we are nevertheless led to conclude that there exists no effective \emph{vibrational} vdW interaction, see Appendix~\ref{app:vibvdW}.

\begin{figure}
    \centering
    \includegraphics[width=1.0\linewidth]{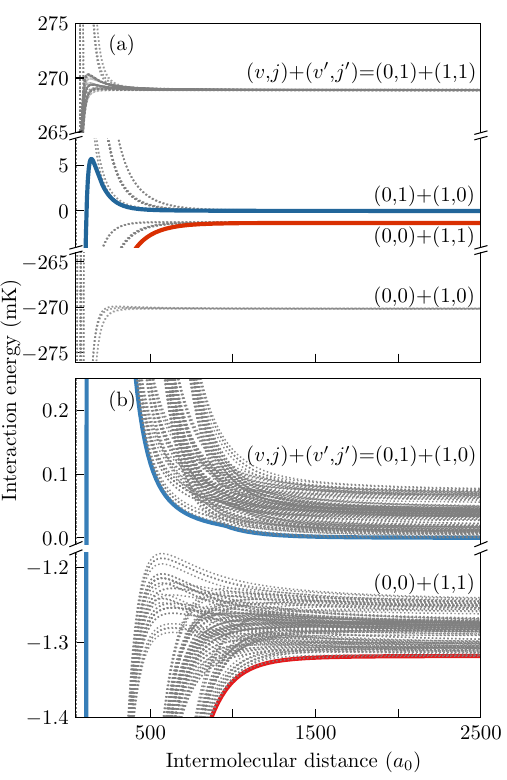}
    \caption{
{\bf Interaction potentials} for NaK molecules interacting by the ro-vibrational vdW interaction.
    Illustrated in gray are adiabatic potential curves for different partial waves $L=0,2,4$,
    and corresponding to different hyperfine states.
    The repulsive (attractive) $s$-wave interaction potentials are shown in blue (red) for the channels $(v,j)+(v',j')=(0,1)+(1,0)$ ($(0,0)+(1,1)$) in the hyperfine ground state.
    Additional thresholds such as $(v,j)+(v',j') = (0,0)+(1,0)$ and $(0,1)+(1,1)$ are shown to occur at different energies on the scale of the rotational constant, and lead to well separated potential curves.
    Panel~(b) provides an expanded view.
    \label{fig:potential}
    }
\end{figure}

\section{Coupled-channels calculations \label{sec:CC}}

We describe collisions between ultracold molecules by coupled-channel calculations that include an absorbing boundary condition at short range, which models collisional loss by sticky collisions.
These calculations quantitatively describe universal collisional loss~\cite{idziaszek2010universal}, and microwave shielding~\cite{karman2018microwave,anderegg:21,schindewolf:22,chen:23,bigagli:23,bigagli:24}.
Most of the details of these calculations have been described in Ref.~\cite{karman:25},
and we here summarize the key aspects and detail the necessary extensions to include the molecular vibration.

A single molecule is described by the Hamiltonian
\begin{align}
    \hat{H}^{(X)} = \hat{H}^{(X)}_0 + \hat{H}^{(X)}_\mathrm{hf} + \hat{H}^{(X)}_\mathrm{Zeeman} + \hat{H}_{\mathrm{ac},\sigma}^{(X)} + \hat{H}_{\mathrm{ac},\pi}^{(X)}.
    \label{eq:monomer}
\end{align}
where the rotation-vibrational Hamiltonian
\begin{align}
 \hat{H}^{(X)}_0 &= \sum_{v,j,m} |v,j,m\rangle\langle v, j, m|\, E(v,j),
\end{align}
is fully specified by the Dunham expansion Eq.~\eqref{eq:dunham}.
In our previous work\cite{karman:25} this was limited to the term $B_\mathrm{rot} \hat{j}^2$.
The remaining terms in the monomer Hamiltonian Eq.~\eqref{eq:monomer} describe the hyperfine structure,
Zeeman interaction with an external magnetic field $B$,
and coupling to microwave electric fields, respectively, see Ref.~\cite{karman:25}.

The total Hamiltonian for the pair of colliding molecules in the center of mass frame contains, in addition to the monomer Hamiltonians discussed above,
relative kinetic energy and an intermolecular interaction potential, see Ref.~\cite{karman:25}.
We approximate the interaction by the longest range contribution, the dipole-dipole interaction
\begin{align}
\hat{V} = -\frac{\sqrt{30}}{4\pi \epsilon_0 R^3} \left[ \left[ \hat{{d}}^{(1)}(\bm{r}_A) \otimes \hat{{d}}^{(1)}(\bm{r}_B)  \right]^{(2)} \otimes C^{(2)}(\hat{R}) \right]^{(0)}_0,
\label{eq:dipdip}
\end{align}
where $C^{(2)}(\hat{R})$ is a rank-two tensor with spherical components given by Racah-normalized spherical harmonics depending on the polar coordinates of $\hat{R}$, the intermolecular vector.
The quantity
\begin{align}
    \left[ \hat{A}^{(k_A)} \otimes \hat{B}^{(k_B)}\right]^{(k)}_q = \sum_{q_A,q_B} \hat{A}^{(k_A)}_{q_A} \hat{B}^{(k_B)}_{q_B} \langle k_A q_A k_B q_B | k q \rangle
\end{align}
indicates a spherical tensor product.
For the $m$ spherical component of the dipole operator we use
\begin{align}
\hat{{d}}_m(\bm{r}_A) = \left[ d_e + \frac{\partial d}{\partial r} \sqrt{\frac{\hbar}{2\mu_A \omega_e}} \left(\hat{a}+\hat{a}^\dagger\right)\right] C_{1,m}(\hat{r}),
\label{eq:dipole}
\end{align}
where $\mu_A$ is the reduced mass for the vibration of molecule $A$,
and $\hat{a} = \sum_v \sqrt{v}\ |v\rangle\langle v+1|$ is the annihilation operator for the molecular vibration.
We note that our previous work~\cite{karman:25} included only the first term in Eq.~\eqref{eq:dipole},
whereas the second term gives the contribution of the vibrational transition dipole moment to the dipolar interaction.

We use a completely uncoupled primitive basis set.
For molecule $X=A,B$ this consists of products of vibrational states $|v\rangle$,
rotational states, $|j, m\rangle$, with position representation
\begin{align}
    \langle \hat{r}^{(X)} | j_x m_x \rangle  = \sqrt{\frac{2j_X+1}{4\pi}} C_{j_X,m_X}(\hat{r}^{(X)}),
\end{align}
and nuclear spin states $|i_1 m_1\rangle|i_2 m_2\rangle$.
Where microwaves are included,
the state of the microwave fields is described in the photon number basis, $|N_\nu\rangle$,
where $N_\nu + N_{0,\nu}$ is the number of photons in field $\nu$ relative to a large reference numbers of photons, $N_0$.

Thus the basis functions describing a single molecule $X$ in the presence of the two microwave fields take the form
\begin{align}
    |v^X\rangle|j^{X} m^{X} \rangle |i^{X}_1 m^{X}_1\rangle|i^{X}_2 m^{X}_2\rangle |N_\sigma\rangle|N_\pi\rangle,
\end{align}
and the matrix elements of the Hamiltonian in this basis can be calculated as described above.
For two molecules in the presence of the two microwave fields, we set up a basis
\begin{widetext}
\begin{align}
    |v^A\rangle|j^{A} m^{A} \rangle |i^{A}_1 m^{A}_1\rangle|i^{A}_2 m^{A}_2\rangle |v^B\rangle |j^{B} m^{B} \rangle |i^{B}_1 m^{B}_1\rangle|i^{B}_2 m^{B}_2\rangle |\ell m_\ell\rangle |N_\sigma\rangle|N_\pi\rangle,
\end{align}
\end{widetext}
which consists of the product of molecule basis sets for each molecule $X=A,B$,
a partial wave basis set $|\ell m_\ell\rangle$ that describes the end-over-end rotation of the two molecules about one another,
and again the Hamiltonians describing both fields.
We note that in this work, we do not actually simultaneously include hyperfine and microwave fields, though we include both in different calculations.
Subsequently, the basis set is adapted to permutation symmetry of identical particles by projecting with
$1\pm\hat{P}$, and appropriately normalizing, where $\hat{P}$ permutes molecules $A$ and $B$~\cite{karman2018microwave}.
The upper (lower) sign applies for identical bosons (fermions).
The total basis is further limited by including functions only for one value of the conserved generalized angular momentum projection
\begin{align}
\mathcal{M} = m^A + m_1^A + m_2^A + m^B + m_1^B + m_2^B + m_\ell + N_\sigma,
\label{eq:mathcalM}
\end{align}
and parity of $\ell$.
Next an asymptotic basis set is determined by numerically diagonalizing the Hamiltonian excluding interaction terms for each value of $\ell$, $m_\ell$.
Required matrices such as the asymptotic Hamiltonian, centrifugal angular momentum, and the interaction are transformed to this permutation-adapted asymptotic representation, in which all scattering calculations are performed.
The basis set is typically truncated by including functions with $j\le 2$, $\ell\le5$, and including no vibrational states excited further higher than the initial state under consideration.

We propagate two linearly independent sets of solutions to the coupled-channels equations using the renormalized Numerov method of Ref.~\cite{janssen2013quantum,karman:23}.
We then impose capture boundary conditions at short distances.
At asymptotically large distances, we impose the usual $S$-matrix boundary conditions corresponding to unit incoming flux in the entrance channel and outgoing flux in all energetically accessible channels.
With these boundary conditions, results are converged with a radial grid ranging from $50$ to $50\,000$ $\alpha_0$ with at least $30$ points per local de Broglie wavelength for NaK.
For KAg, a radial grid extend from $1\,139$ to $10^{7}$ $\alpha_0$, with a minimum of $30$ grid points per local de Broglie wavelength is used.
From these $S$-matrix we compute elastic and inelastic cross sections, see Ref.~\cite{karman:25}

Thermal rate coefficients are calculated by averaging these cross sections over the Maxwell-Boltzmann distribution for a given temperature, where we perform the integration using 19 logarithmically spaced collision energies between $10^{-10}$ and $10^{-5}$ K.

Low-energy scattering can be characterized by the $s$-wave scattering length, $a_s$,
which can be extracted from the $S$-matrix as
\begin{align}
    a_s = \lim_{E\rightarrow 0}\ \frac{1-S_{i,0,0;\ i,0,0}(E)}{ik\left[1+S_{i,0,0;\ i,0,0}\left(E\right)\ \right]},
\end{align}
where $S_{i,0,0;\ i,0,0}$ is the $S$-matrix diagonal element corresponding to the $s$-wave initial channel.
The $S$-matrix is obtained from the coupled-channels calculations described above.
We numerically confirm that the scattering length is energy independent at the lowest energies, typically in the pK to nK range.

\section{Collision rates \label{sec:results}}

The goal of our coupled-channel calculations is to quantitatively put these interaction mechanisms to the test. Specifically, we aim to demonstrate that the ro-vibrational vdW interaction drastically suppresses inelastic collisions,
and boost elastic collisions.
%and that this suppression depends critically on the specific choice of states. 
To this end, we evaluate the loss rates for collisions of two fermionic NaK molecules as a representative system.

\begin{figure}
    \centering
    \includegraphics[width=0.95\linewidth]{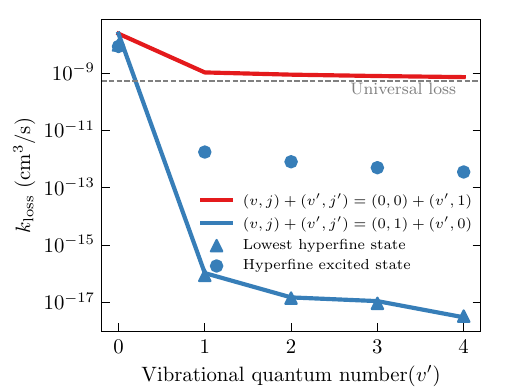}
    \caption{{\bf Rotation-vibration state dependence} of the loss rate for fermionic NaK at 1~nK and a magnetic field $B$ of 500~gauss.
    At this $B$ field,
    the lowest hyperfine is $m_{I_\mathrm{Na}}=3/2$, $m_{I_\mathrm{K}}=-4$ and $M_F = -5/2$ or $-3/2$ for $j=0$ or $1$, respectively.
    The excited hyperfine state shown is $|m_{I_\mathrm{Na}}, m_{I_\mathrm{K}}, m_F\rangle=|3/2,-3,-3/2\rangle$ for $j=0$,
    and $|3/2,-4,-7/2\rangle$ for $j=1$.
    } 
    \label{fig:vibloss}
\end{figure}

The resulting inelastic collision rates at a temperature of 1~nK are shown in Fig,~\ref{fig:vibloss}.
One molecule is prepared in $v=0$ while the second is prepared in $v'$ as indicated by the horizontal axis.
If the $v=0$ molecule is in rotational $j=0$, and the $v'>0$ molecule is in $j=1$,
the ro-vibrational van der Waals interaction is attractive.
This results in universal short-range loss at a rate proportional to $R_6 \propto C_6^{1/4}$,
such that the resulting rate coefficient decreases only slowly with $v'$, which scales the energy defect.
In case the $v=0$ molecule is prepared in rotational state $j=1$, and the $v'>0$ molecule is prepared in $j=0$, the situation is reversed, and the ro-vibrational van der Waals interaction is repulsive.
This suppresses collisional loss by some seven orders of magnitude.
In both cases, if $v'=0$, the interactions are instead given by resonant dipole-dipole couplings, which are longer-ranged and lead to substantially faster collisional loss.

%hyperfine.
When repulsive ro-vibrational vdW interactions suppress collisional loss,
we find that the residual loss rate is sensitive to the hyperfine state,
shown as the markers in Fig.~\ref{fig:vibloss}.
When the molecules are prepared in an excited hyperfine state,
the loss rate coefficient is orders of magnitude higher than that obtained in the hyperfine-free calculation, shown as the solid lines.
We verify that this additional loss is entirely attributed to hyperfine relaxation, that is, inelastic transitions to lower hyperfine levels within the same $(v,j) + (v' j')$ manifold.
However, when the molecules are prepared in their hyperfine ground state,
this relaxation is energetically forbidden, and the loss rate coefficient is suppressed essentially to the hyperfine-free level.

\begin{figure}
    \centering
    \includegraphics[width=0.9\linewidth]{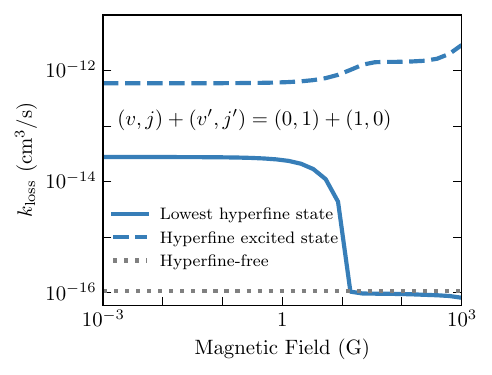}
    \caption{{\bf Magnetic field dependence of collisional loss} rate for fermionic NaK molecules in the lowest and excited hyperfine states of $(v,j)+(v',j')= (0,1)+(1,0)$ as a function of magnetic field at 1~nK. Near $B=20$~G, the loss rate for the lowest hyperfine state exhibits a suppression of approximately two orders of magnitude, while the excited hyperfine state shows a slight increase.
    }
    \label{fig:bfield}
\end{figure}

The results shown in Fig.~\ref{fig:vibloss} are for a magnetic field $B=500$~G.
The full magnetic field dependence is shown in Fig.~\ref{fig:bfield}. At low fields, the loss rates for both the lowest and excited hyperfine states are comparable ($10^{-14}-10^{-12}$ cm$^{3}$/s). As the magnetic field approaches $B=20$~G, the loss rate for the lowest hyperfine state is suppressed by two orders of magnitude, dropping to the hyperfine-free result. Above this threshold $B$ value, the ground-state loss rate is essentially constant, demonstrating that a small magnetic field of roughly $20$ G is sufficient to suppress residual loss due to hyperfine relaxation. The loss rate for the excited hyperfine state does not experience this suppression and instead shows a slight increase with the field. 

\section{Universality \label{sec:universality}}

\begin{figure}
    \centering
    \includegraphics[width=0.9\linewidth]{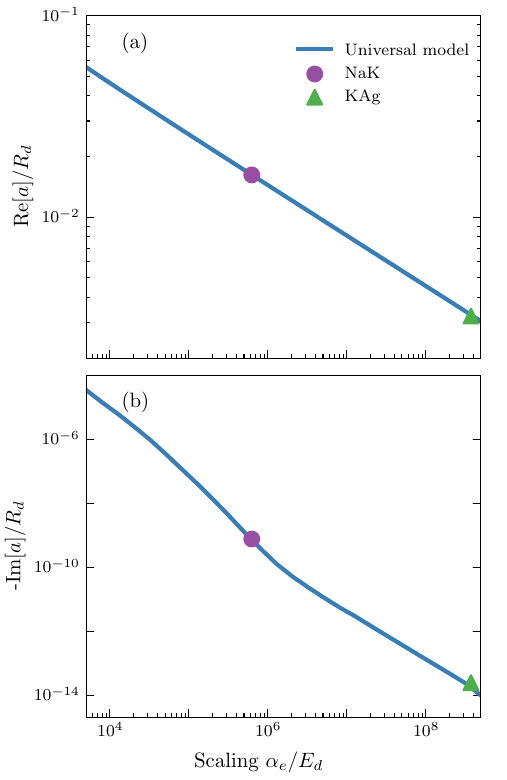}
    \caption{{\bf Universal scaling of the scattering length} for ro-vibrational van der Waals repulsion.
    The universal relation is calculated in a minimal model that only accounts for the pair states $(v,j)+(v',j') = (0,1)+(1,0)$ to $(0,0)+(1,1)$, split by $2\alpha_e$, see main text.
    Panels (a) and (b) shows the real and imaginary part of the scattering length, respectively.
    The scattering length is expressed in dipolar length units,
    and shown as a universal function of the dimensionless parameter $\alpha_e/E_d$.
    Markers indicate the results of full coupled-channels calculations for NaK and KAg,
    and the agreement with the minimal model verifies the universality.
    }
    \label{fig:universality}
\end{figure}

Next, we consider the universality of the ro-vibrational van der Waals interaction for linear diatomic molecules.
The essence of the repulsive interaction is the dipole-dipole coupling between pairs of molecule states $(v,j)+(v',j') = (0,1)+(1,0)$ to $(0,0)+(1,1)$.
The dipolar interaction is characterized by a dipolar length $R_d =  md^2/8\pi\epsilon_0\hbar^2$ and energy scale $E_d=\hbar^2/mR_d^2$,
and the energy defect $2\alpha_e$ sets an additional energy and associated length scale.
The zero-energy collision properties, characterized by the scattering length, $a_s$,
do not depend on any other length or energy scales,
whereas at finite collision energy, the collision energy itself and the associated de Broglie wavelength set an additional energy and length scale.
Hence, we expect that $a_s/R_d$ is a universal function of the ratio $\alpha_e/E_d$,
meaning that $a_s/R_d$ can depend \emph{only} on the ratio $\alpha_e/E_d$, but it cannot depend on molecular constants in any other way.
Another way of phrasing this is that different combinations of molecular parameters (mass, dipole moment, or rotation-vibration coupling constant) that lead to the same ratio $\alpha_e/E_d$, lead to the same $a_s/R_d$.

This is perhaps clearest for the elastic properties,
for which we can ignore the unlikely non-adiabatic transitions that correspond to inelastic collisions,
and consider collisions on the upper repulsive adiabatic potential.
This is essentially a repulsive van der Waals potential,
for which\cite{kevin}
\begin{align}
    {a_s}/{R_6} = -\frac{\Gamma(-\frac{1}{4})}{2\Gamma(\frac{1}{4})} \approx 0.676.
\end{align}
Using $C_6 = d_e^4/9\alpha_e$, $R_d/R_6 = (9\alpha_e / 4 E_d)^{1/4}$, this gives
\begin{align}
a_s/R_d \approx 0.552 \left(\frac{\alpha_e}{E_d}\right)^{-1/4}.
\end{align}
For the imaginary part of the scattering length, we determine the dependence on $\alpha_e/E_d$ numerically by performing coupled channels calculations limited to the two essential pairs of molecule states $(v,j)+(v',j') = (0,1)+(1,0)$ to $(0,0)+(1,1)$.
Here, we find a scaling close to the power-law $-\mathrm{Im}[a_s]/R_d \propto (\alpha_e/E_d)^{-2}$.
The numerical results for both the real and imaginary parts of the scattering length are shown in Fig.~\ref{fig:universality}.

Real molecules have additional energy levels, in addition to the essential pairs of molecule states $(v,j)+(v',j') = (0,1)+(1,0)$ to $(0,0)+(1,1)$, considered above.
These energy levels are separated from these channels by an amount that depends not only on $\alpha_e$,
but also on the rotational constant $B_e$, for example.
These additional energy scales could in principle break the universality.
Here, we verify numerically that for representative molecules, these additional energy scales play no role and the simple universality discussed above holds.
To this end we performed coupled-channels calculations that include the full ro-vibrational energy structure of NaK and KAg molecules.
The resulting scattering lengths, shown as markers in Fig.~\ref{fig:universality}, agree closely with the universal result.

We note that, in order to simplify the discussion above, we have ignored the existence of a second dipolar length and energy scale, $R_d'$ and $E_d'$, set by the vibrational transition dipole moment.
This leads to first-order dipolar interactions in each of the essential channels of the molecular pair $(v,j)+(v',j') = (0,1)+(1,0)$ and $(0,0)+(1,1)$.
The existence of this interaction energy scale --like the molecules internal energy level structure-- in principle could break universality.
However, we neglected this interaction in our universal minimal model,
but included it in our full coupled-channels calculations,
such that the agreement between the two establishes that the vibrational off-diagonal dipolar interaction plays no role for representative ultracold molecules.
Finally, the hyperfine structure of real molecules sets further energy scales that are not included in the present discussion,
but we have already demonstrated in Sec.~\ref{sec:results} that hyperfine plays no role for molecules prepared in their hyperfine ground states.

From this universality we can understand the effectiveness of the ro-vibrational vdW interaction and its dependence on molecular parameters.
Larger $\alpha_e$ suppresses non-adiabatic transitions from the upper repulsive potential to the lower attractive one,
which in the universal two-level description is the only loss channel.
The inelastic part of the scattering length is significantly suppressed as $-\mathrm{Im}[a_s] \propto \alpha_e^{-2}$.
Increasing $\alpha_e$ also results in a weak suppression of the elastic part of the scattering length $\mathrm{Re}[a_s] \propto \alpha_e^{-1/4}$.
Since the imaginary part of the scattering length is suppressed much more strongly than the real part,
the ro-vibrational vdW interaction can be considered more effective for larger $\alpha_e$.
At fixed $\alpha_e$, we have the scaling $\mathrm{Re}[a_s] \propto R_d^{1/2}$ while $\mathrm{Im}[a_s] \propto R_d^{-3}$.
This implies that when going to more dipolar species, at fixed $\alpha_e$,
the elastic cross section grows while the inelastic cross section is suppressed further.
As shown in Fig.~\ref{fig:universality}, these effects result in ro-vibrational vdW interactions that become orders of magnitude more effective when transitioning from NaK to the \emph{ultrapolar} KAg.

\section{Direct evaporation of a Fermi Mixture \label{sec:evap}}

\begin{figure}
    \centering
    \includegraphics[width=1.0\linewidth]{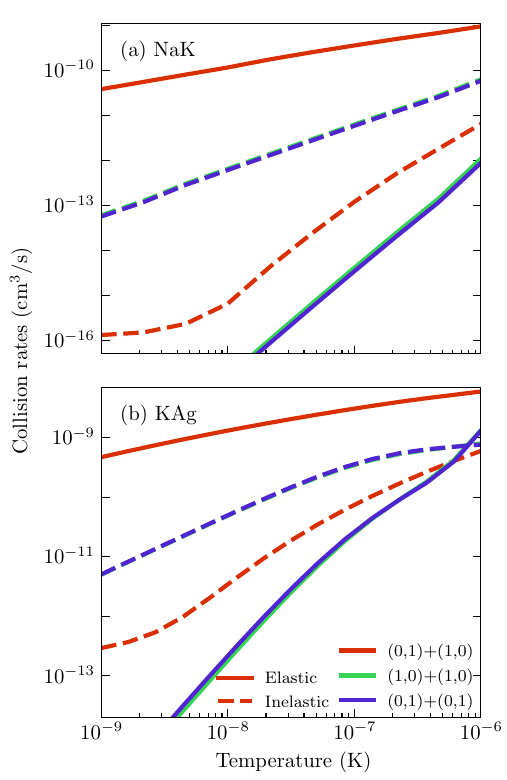}
    \caption{{\bf Collision rates in a Fermi mixture} for various pairs of rotational-vibrational state of (a) NaK and (b) KAg as a function of temperature.
}
    \label{fig:fermion}
\end{figure}

Cooling down to a degenerate quantum gas of fermions has so far been achieved by assembly from deeply degenerate atoms~\cite{demarco:19,duda:23},
and by evaporative cooling~\cite{valtolina:20,matsuda:20,li:21,anderegg:21,schindewolf:22,bigagli:24,shi2025boseeinsteincondensateultracoldsodiumrubidium}.
Efficient evaporative cooling requires a high ``$\gamma$ ratio'' of elastic to inelastic collision rates.
For molecules, this has generally required the suppression of collisional loss by shielding,
where external fields are used to engineer repulsive long-range interactions between molecules.
Here, we consider direct evaporation without active collisional shielding using external fields by using a mixture of fermionic molecules in $v,j=0,1$ and $v',j'=1,0$ states.

For a pair of molecules in the same internal state,
the dominant interaction is the rotational van der Waals interaction,
which results in universal loss~\cite{idziaszek2010universal}.
For fermionic molecules, these $p$-wave collisions lead to elastic and inelastic cross sections that are suppressed at low temperature as $T^2$ and $T$, respectively.
For a pair of molecules in different internal states,
the dominant interaction is the ro-vibrational van der Waals interaction.
These collisions can proceed by $s$-wave,
leading to $T^{1/2}$ and $T^{0}$ scaling of the elastic and inelastic collision rates.
In absolute terms, the elastic collision rate which is proportional to $R_6^2\propto C_6^{1/2}$ is enhanced by orders of magnitude for the stronger ro-vibrational interaction, when compared to the rotational one.
The inelastic rate is suppressed by orders of magnitude due to the repulsive ro-vibrational vdW interaction.

Figure~\ref{fig:fermion}(a) shows the relevant collision rates quantitatively for NaK molecules at temperatures between 1~nK and 1~$\mu$K.
The rates for inelastic $s$-wave collisions between molecules in different internal states and for elastic $p$-wave collisions between molecules in identical internal states are so far suppressed that they play essentially no role.
The elastic $s$-wave rate dominates over the inelastic $p$-wave rate by a factor of 15 at $T=1~\mu$K,
and this ``$\gamma$ ratio'' of elastic-to-inelastic collisions improves substantially towards lower temperature in accordance with the Wigner threshold laws.

Figure~\ref{fig:fermion}(b) shows the relevant collision rates for KAg molecules.
Based on the universality discussed in Sec.~\ref{sec:universality},
we expect the elastic $s$-wave collision rate to scale as $R_d (k_BT/\alpha_e)^{1/2}$ for the ro-vibrational van der Waals interaction,
while the $p$-wave inelastic rate scales as $R_d^{3/2}\ k_BT\ B_e^{-3/4}$ for the rotational van der Waals interaction \cite{idziaszek2010universal}.
Therefore, the $\gamma$ ratio at fixed temperature is expected to scale as $R_d^{-1/2}$,
leading to a small decrease for the ultrapolar molecules.
The reduction of $\alpha_e$ and $B_e$ for KAg compared to NaK lead to a small increase and decrease of the $\gamma$ ratio, respectively.
In general, the scaling of the $\gamma$ ratio with molecular parameters is not very steep.
The more qualitative difference between the two cases is the onset of threshold scaling of the collision rates,
which occurs at lower temperature for KAg due to the lower dipolar energy scale.

In general we can conclude the ro-vibrational vdW interaction can enable direct evaporation of Fermi mixtures of molecules in different vibrational states.
The scaling of the $\gamma$ ratio at fixed absolute temperature $(R_d\,\alpha_e)^{-1/2}$ indicates the ratio is better for smaller vibration rotation coupling constant $\alpha$, which boosts the elastic cross section, 
and for \emph{less} dipolar molecules, because a stronger dipolar interaction boosts the inelastic collision rate slightly more than the elastic one.
However, these scalings scalings are not steep enough to create order of magnitude performance differences between different molecules.
The $\gamma$ ratio is thus expected to be roughly on the order of tens around microkelvin temperatures,
and to further improve at lower temperature, for many molecules.
This $\gamma$ ratio is not as comfortably high as what can be achieved by active collisional shielding~\cite{schindewolf:22,bigagli:24,shi2025boseeinsteincondensateultracoldsodiumrubidium},
but it could enable evaporative cooling simply by creating a mixture of different vibrational states without active external field control of the collisions.

\section{Discussion \label{sec:discussion}}

We believe that the ro-vibrational van der Waals interactions described in this work will enable new strategies for controlling collisions and interactions between ultracold molecules,
leading a multitude of powerful applications in quantum simulation and many-body physics.
In addition to enabling direct evaporation of Fermi mixtures discussed above, in Sec.~\ref{sec:evap},
we briefly discuss some opportunities here.

\begin{figure}
    \centering
    \includegraphics[width=0.9\linewidth]{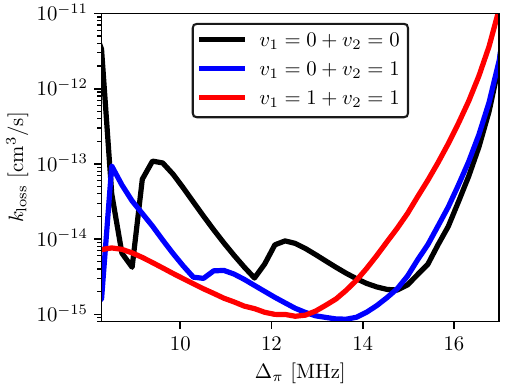}
    \caption{{\bf Simultaneous double microwave shielding} of ground and excited vibrational states. Two-body loss rates, $k_{\mathrm{loss}}$, as a function of the detuning of the $\pi$-polarized microwave field, $\Delta_{\pi}$. Calculations performed for NaCs at $\Omega_{\sigma} = 44.7\times 2\pi\,\mathrm{MHz} ,\,\Omega_{\pi} = \Delta_{\sigma} = 22.4\times 2\pi\,\mathrm{MHz}$.
    }
    \label{fig:vib_shielding}
\end{figure}

%Compatibility with shielding 
The ro-vibrational van der Waals interactions introduced in this work can be combined with active collisional shielding techniques, specifically double microwave shielding~\cite{bigagli:24, karman:25}. Double microwave shielding utilizes $\sigma^{+}$ and $\pi$-polarized microwave fields detuned from the $j=0 \to 1$ rotational transition, to engineer a repulsive long-range barrier that suppresses collisional losses for molecules in their ground vibrational state. 
These microwave fields can simultaneously shield molecules occupying different vibrational states such as $v=0$ and $v=1$.
Because of the rotation-vibration coupling, the $j=0 \to 1$ transition energy in the vibrationally excited $v=1$ state is smaller than that of the $v=0$ ground state by $2\alpha_{e}$,
which is on the order of MHz and therefore comparable to typical Rabi frequencies and detunings used in double microwave shielding.
Consequently, the two microwave fields dressing the $v=0$ molecules act simultaneously on the $v=1$ manifold, simply with an effective detuning shifted by $+2\alpha_{e}$.
It is possible to find microwave configurations ---the two Rabi frequencies, $\Omega_{\sigma}, \Omega_{\pi}$, and the two detunings, $\Delta_{\sigma}, \Delta_{\pi}$--- that simultaneously suppress two-body loss rates for collisions involving molecules in both vibrational states.
As an example, Fig. \ref{fig:vib_shielding} shows computed two-body loss rates for the NaCs molecule in a specific microwave field configuration, demonstrating the suppression of losses for all identical ($v=0 + v=0$ and $v=1 + v=1$) and distinguishable ($v=0 + v=1$) collisions.

The ability to simultaneously shield molecules in different vibrational states opens up exciting avenues for quantum simulation and spin physics. By preparing a stable mixture of vibrationally ground and excited molecules, one can encode a pseudo-spin $1/2$ system where the field-dressed $v=0$ and $v=1$ states serve as the effective spin quantum numbers. 
Double microwave shielding then ensures not only the collisional stability of the gas, but also provides tunability of the effective interactions between these pseudo-spin states via the induced dipolar interactions. 
Specifically, this enables realization of a tunable dipolar density-density, spin-density, and Ising exchange interactions in a collisionally stable gas.
This provides a robust platform for exploring quantum simulation and many-body physics with state-dependent long-range interactions, as detailed in the accompanying work~\cite{spin}.

The highly-tunable state-dependent interactions are also interesting in the polaron setting,
where we envision creating impurities of molecules in $v=1$ in a bath of $v=0$ molecules, for example.
Double microwave shielding can simultaneously stabilize the bath and impurity with respect to collisions.
It is possible to tune the microwave parameters such that the dipolar interaction between $v=0$ bath molecules switched off, or compensated.
Because the effective detuning for $v=1$ is different, however, dipolar interactions between $v=1$ impurity molecules and $v=0$ bath molecules are not compensated by the same microwave fields.
This realizes counter-intuitive ``asymmetric'' interactions\cite{saffman2009efficient};
The $v=0$ bath molecules are effectively non-polar, but they have dipole-dipole interactions with $v=1$ impurity molecules.
That is, the dipole moment of the $v=0$ molecules appears to depend on their interaction partner.
Especially for the Fermi polaron this setting is interesting as it enables realizing a strongly interacting dipolar polaron with a non-interacting bath.
By tuning away from compensation dipolar interactions can be re-introduced in the bath,
which can lead to unconventional $p$ superfluidity,
and this new platform enables studying the competition between these various pairing mechanisms.
This setup is explored in the future work~\cite{polaron}.

We also envision that the ro-vibrational vdW interaction may be a powerful tool to realize deterministic loading of tweezers with ultracold molecules, using schemes similar to that in Ref.~\cite{walraven:24b}.
The general strategy there was repeated loading of molecules into conservative trapping potentials using laser cooling.
Successfully loaded molecules are ``shelved'' in a rotationally excited state,
such that the shelved molecule is shielded from collisions with subsequently loaded molecules by the rotational vdW interaction.
A blockade is implemented to make sure only one molecule per tweezer can be shelved,
leading to high-probability loading of single molecules, realizing highly scalable tweezer arrays.
In this scheme, the probability of loading single molecules is limited to $\sim80~\%$ by residual collisional loss of molecules interacting by rotational vdW repulsion.
This limiting loading fidelity is comparable to what is possible by controlling light assisted collisions between atoms~\cite{kaufman:21}.
Since the ro-vibrational vdW interaction can be orders of magnitude more effective at suppressing collisional loss,
this novel interaction can substantially improve deterministic loading of molecules~\cite{tweezer}.

The ro-vibrational vdW interaction may also enable infrared shielding,
where molecules are prepared in identical internal states, e.g. $v,j=1,0$,
and are dressed on the ro-vibrational transition $v,j=1,0 \longleftrightarrow 0,1$.
As a result, a pair of molecules in states $(v,j)+(v',j')=(1,0)+(1,0)$ is dressed with $(1,0)+(0,1)$ and $(0,1)+(0,1)$.
This dressing mixes in strongly repulsive ro-vibrational vdW interactions that could realize collisional shielding.
The difficulty is that for many molecules, the vibrational transition occurs in the terahertz regime where it is difficult to realize high power.
However, there are exceptions as for lighter molecules the vibrational transition are shifted further into the infrared.
For example, for NH molecules the vibrational transition is shifted to $3~\mu$m~\cite{radford1975imine,wayne1976laser}.
The lifetime of vibrationally excited states is tens of miliseconds,
so that off-resonant dressing may be possible with one-body lifetimes approaching the second scale.
Whether this IR shielding can be effective is not immediately clear as $\alpha_e \simeq 19$~GHz is orders of magnitude larger than is typical for the heavier assembled molecules.
This should lead to a weaker ro-vibrational vdW repulsion, with $C_6^\mathrm{ro-vib} = d_e^4/9\alpha_e = 5\,000$ a.u., but on the other hand the competing interactions are also weaker for this molecule.
For example, the rotational vdW interaction $C_6^\mathrm{rot} = d_e^4/6B_e = 280$~a.u.\ and the electronic vdW interaction $C_6^\mathrm{elec} = 47$~a.u.~\cite{koput2015ab}.
Hence, the ro-vibrational vdW interaction is still the dominant interaction and future research may reveal whether this can realize effective IR shielding of NH molecules.

More broadly, the ro-vibrational vdW interaction discovered here is a novel resource for interaction control in different internal states.
A setting in which this interaction is naturally accessed is quantum simulation with synthetic dimensions\cite{sundar2018synthetic,feng2022quantum} encoded in ro-vibrational degrees of freedom.
Ro-vibrational vdW repulsion can also be a powerful tool for quantum simulation with fermionic molecules in optical lattices,
where tunneling of two molecules onto the same lattice site can otherwise lead to collisional loss.
Together with the applications of Fermi mixture evaporation, tunably interacting spin mixtures~\cite{spin,polaron}, enhanced tweezer loading~\cite{tweezer}, and direct infrared shielding, discussed above,
we conclude that the ro-vibrational vdW interaction can be a powerful resource with many potential applications for quantum science and many-body physics with ultracold molecules.

\section{Acknowledgement}

We thank Edvardas Narevicius, Sebastian Will, Ian Stevenson, Eugen Dizer, Arthur Christianen and Richard Schmidt for useful discussions.
We thank Jacek Koput, Luk{\'a}{\v{s}} Pa{\v{s}}teka and Anastasia Borschevsky for sharing ab initio calculations.
This work was supported by NWO VIDI (grant ID 10.61686/AKJWK33335).
The research was funded by the European Union (Project No.~101269084, HORIZON-MSCA-2025-PF, 2STICKY). The views and opinions expressed are, however, those of the authors only
and do not necessarily reflect those of the European Union or the European Research Executive Agency. Neither
the European Union nor the granting authority can be held responsible for them.

\appendix

\section{Vibrational vdW \label{app:vibvdW}}

%vibrational vdW
For many molecules, the vibrational transition dipole moment is smaller than for rotational transitions,
and the energy defect on the order of $\omega_e$ is orders of magnitude larger than rotational excitation energies.
This results in a vibrational vdW interaction that is orders of magnitude weaker than rotational vdW.

There is an exception to this, which is for transitions of the type $v,v' \rightarrow v-1,v'+1$,
where the energy denominator is on the order of the anharmonicity $\omega_ex_e$.
The anharmonicity $\omega_ex_e$ is typically only a few times larger than the rotational constant, $B_e$,
and both are on the GHz scale for many molecules that we consider, see Table.~\ref{tab:constants}.
Therefore, it is even possible to find transitions for which the rotational excitation and vibrational de-excitation energies can cancel to a large extent.
For example, the transition $(v,j)+(v,j) = (1,2)+(1,2) \rightarrow (0,3)+(2,3)$ has an energy denominator that is $2\omega_ex_e-12B_e+18\alpha_e$,
which for most molecules considered is only a few percent of the anharmonicity.
For KAg, for example, this energy denominator is $\sim40~\%$ of the rotational constant,
i.e. an order of magnitude smaller than the $4B_e$ energy denominator for rotational vdW for molecules in their ground state.
What is interesting about this, apart from the small energy denominator, is that this can lead to repulsive vdW interactions between two molecules in the same internal state, here $v,j=1,2$, such that this could be applied to a bulk gas of molecules in the same internal state. 

\begin{figure}
    \centering
    \includegraphics[width=0.9\linewidth]{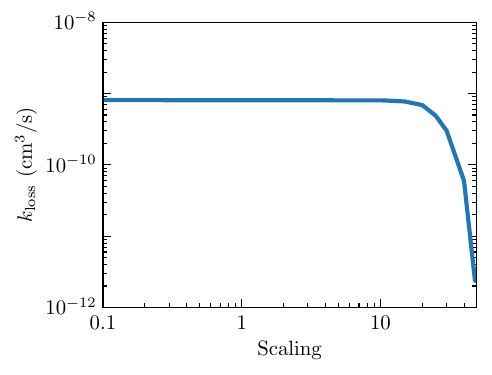}
    \caption{{\bf Vibrational vdW Interaction.} Collisional loss rate for KAg molecules in $v,j=1,2$ as a function of artificial scaling of the vibrational transition dipole moment.
    When the vibrational transition dipole moment is scaled by a factor $\sim50$, it is made comparable to the transtition dipole for purely rotational transitions.
    In this case, virtual de-excitation $(v,j)+(v',j')=(1,2)+(1,2) \rightarrow (0,3)+(2,3)$ suppresses loss by a repulsive vibrational vdW interaction.
    For any real molecule, however, the vibrational transition dipole moment is weaker than the rotational one by more than an order of magnitude,
    and we conclude that this vibrational vdW cannot be effective.
    }
    \label{fig:vibvdW}
\end{figure}

Despite it being possible to realize order-of-magnitude smaller energy denominators when compared to purely rotational transitions,
the vibrational transition dipole moments are one to several orders of magnitude smaller than the rotational ones,
depending strongly on the molecular species.
Due to the steep scaling of the $C_6\propto d_e^4/\Delta E$ coefficient with transition dipole moment,
this nevertheless makes the vibrational vdW interaction generally ineffective compared to the rotational one.
We demonstrate this in Fig~\ref{fig:vibvdW} by showing that effective shielding by this vibrational vdW interaction for KAg molecules in $v,j=1,2$ requires artificially increasing the vibrational transition dipole moment by a factor $\sim 50$.
Therefore we conclude there exists no effective vibrational vdW interaction between ultracold polar molecules.

\bibliography{bibliography}

\begin{thebibliography}{96}%
\makeatletter
\providecommand \@ifxundefined [1]{%
 \@ifx{#1\undefined}
}%
\providecommand \@ifnum [1]{%
 \ifnum #1\expandafter \@firstoftwo
 \else \expandafter \@secondoftwo
 \fi
}%
\providecommand \@ifx [1]{%
 \ifx #1\expandafter \@firstoftwo
 \else \expandafter \@secondoftwo
 \fi
}%
\providecommand \natexlab [1]{#1}%
\providecommand \enquote  [1]{``#1''}%
\providecommand \bibnamefont  [1]{#1}%
\providecommand \bibfnamefont [1]{#1}%
\providecommand \citenamefont [1]{#1}%
\providecommand \href@noop [0]{\@secondoftwo}%
\providecommand \href [0]{\begingroup \@sanitize@url \@href}%
\providecommand \@href[1]{\@@startlink{#1}\@@href}%
\providecommand \@@href[1]{\endgroup#1\@@endlink}%
\providecommand \@sanitize@url [0]{\catcode `\\12\catcode `\$12\catcode
  `\&12\catcode `\#12\catcode `\^12\catcode `\_12\catcode `\%12\relax}%
\providecommand \@@startlink[1]{}%
\providecommand \@@endlink[0]{}%
\providecommand \url  [0]{\begingroup\@sanitize@url \@url }%
\providecommand \@url [1]{\endgroup\@href {#1}{\urlprefix }}%
\providecommand \urlprefix  [0]{URL }%
\providecommand \Eprint [0]{\href }%
\providecommand \doibase [0]{https://doi.org/}%
\providecommand \selectlanguage [0]{\@gobble}%
\providecommand \bibinfo  [0]{\@secondoftwo}%
\providecommand \bibfield  [0]{\@secondoftwo}%
\providecommand \translation [1]{[#1]}%
\providecommand \BibitemOpen [0]{}%
\providecommand \bibitemStop [0]{}%
\providecommand \bibitemNoStop [0]{.\EOS\space}%
\providecommand \EOS [0]{\spacefactor3000\relax}%
\providecommand \BibitemShut  [1]{\csname bibitem#1\endcsname}%
\let\auto@bib@innerbib\@empty
%</preamble>
\bibitem [{\citenamefont {Karman}\ \emph {et~al.}(2024)\citenamefont {Karman},
  \citenamefont {Tomza},\ and\ \citenamefont
  {P{\'e}rez-R{\'\i}os}}]{karman:24}%
  \BibitemOpen
  \bibfield  {author} {\bibinfo {author} {\bibfnamefont {T.}~\bibnamefont
  {Karman}}, \bibinfo {author} {\bibfnamefont {M.}~\bibnamefont {Tomza}},\ and\
  \bibinfo {author} {\bibfnamefont {J.}~\bibnamefont {P{\'e}rez-R{\'\i}os}},\
  }\href {https://doi.org/10.1038/s41567-024-02467-3} {\bibfield  {journal}
  {\bibinfo  {journal} {Nature Phys.}\ }\textbf {\bibinfo {volume} {20}},\
  \bibinfo {pages} {722} (\bibinfo {year} {2024})}\BibitemShut {NoStop}%
\bibitem [{\citenamefont {Cooper}\ and\ \citenamefont
  {Shlyapnikov}(2009)}]{cooper:09}%
  \BibitemOpen
  \bibfield  {author} {\bibinfo {author} {\bibfnamefont {N.~R.}\ \bibnamefont
  {Cooper}}\ and\ \bibinfo {author} {\bibfnamefont {G.~V.}\ \bibnamefont
  {Shlyapnikov}},\ }\href {https://doi.org/10.1103/PhysRevLett.103.155302}
  {\bibfield  {journal} {\bibinfo  {journal} {Phys. Rev. Lett.}\ }\textbf
  {\bibinfo {volume} {103}},\ \bibinfo {pages} {155302} (\bibinfo {year}
  {2009})}\BibitemShut {NoStop}%
\bibitem [{\citenamefont {Gorshkov}\ \emph {et~al.}(2011)\citenamefont
  {Gorshkov}, \citenamefont {Manmana}, \citenamefont {Chen}, \citenamefont
  {Ye}, \citenamefont {Demler}, \citenamefont {Lukin},\ and\ \citenamefont
  {Rey}}]{gorshkov2011}%
  \BibitemOpen
  \bibfield  {author} {\bibinfo {author} {\bibfnamefont {A.~V.}\ \bibnamefont
  {Gorshkov}}, \bibinfo {author} {\bibfnamefont {S.~R.}\ \bibnamefont
  {Manmana}}, \bibinfo {author} {\bibfnamefont {G.}~\bibnamefont {Chen}},
  \bibinfo {author} {\bibfnamefont {J.}~\bibnamefont {Ye}}, \bibinfo {author}
  {\bibfnamefont {E.}~\bibnamefont {Demler}}, \bibinfo {author} {\bibfnamefont
  {M.~D.}\ \bibnamefont {Lukin}},\ and\ \bibinfo {author} {\bibfnamefont
  {A.~M.}\ \bibnamefont {Rey}},\ }\href
  {https://doi.org/10.1103/PhysRevLett.107.115301} {\bibfield  {journal}
  {\bibinfo  {journal} {Phys. Rev. Lett.}\ }\textbf {\bibinfo {volume} {107}},\
  \bibinfo {pages} {115301} (\bibinfo {year} {2011})}\BibitemShut {NoStop}%
\bibitem [{\citenamefont {Micheli}\ \emph {et~al.}(2006)\citenamefont
  {Micheli}, \citenamefont {Brennen},\ and\ \citenamefont
  {Zoller}}]{micheli2006toolbox}%
  \BibitemOpen
  \bibfield  {author} {\bibinfo {author} {\bibfnamefont {A.}~\bibnamefont
  {Micheli}}, \bibinfo {author} {\bibfnamefont {G.~K.}\ \bibnamefont
  {Brennen}},\ and\ \bibinfo {author} {\bibfnamefont {P.}~\bibnamefont
  {Zoller}},\ }\href@noop {} {\bibfield  {journal} {\bibinfo  {journal} {Nature
  Physics}\ }\textbf {\bibinfo {volume} {2}},\ \bibinfo {pages} {341} (\bibinfo
  {year} {2006})}\BibitemShut {NoStop}%
\bibitem [{\citenamefont {Cornish}\ \emph {et~al.}(2024)\citenamefont
  {Cornish}, \citenamefont {Tarbutt},\ and\ \citenamefont
  {Hazzard}}]{cornish:24}%
  \BibitemOpen
  \bibfield  {author} {\bibinfo {author} {\bibfnamefont {S.~L.}\ \bibnamefont
  {Cornish}}, \bibinfo {author} {\bibfnamefont {M.~R.}\ \bibnamefont
  {Tarbutt}},\ and\ \bibinfo {author} {\bibfnamefont {K.~R.}\ \bibnamefont
  {Hazzard}},\ }\href@noop {} {\bibfield  {journal} {\bibinfo  {journal}
  {Nature Physics}\ }\textbf {\bibinfo {volume} {20}},\ \bibinfo {pages} {730}
  (\bibinfo {year} {2024})}\BibitemShut {NoStop}%
\bibitem [{\citenamefont {DeMille}(2002)}]{demille2002quantum}%
  \BibitemOpen
  \bibfield  {author} {\bibinfo {author} {\bibfnamefont {D.}~\bibnamefont
  {DeMille}},\ }\href@noop {} {\bibfield  {journal} {\bibinfo  {journal} {Phys.
  Rev. Lett.}\ }\textbf {\bibinfo {volume} {88}},\ \bibinfo {pages} {067901}
  (\bibinfo {year} {2002})}\BibitemShut {NoStop}%
\bibitem [{\citenamefont {Ruttley}\ \emph {et~al.}(2025)\citenamefont
  {Ruttley}, \citenamefont {Hepworth}, \citenamefont {Guttridge},\ and\
  \citenamefont {Cornish}}]{ruttley2025long}%
  \BibitemOpen
  \bibfield  {author} {\bibinfo {author} {\bibfnamefont {D.~K.}\ \bibnamefont
  {Ruttley}}, \bibinfo {author} {\bibfnamefont {T.~R.}\ \bibnamefont
  {Hepworth}}, \bibinfo {author} {\bibfnamefont {A.}~\bibnamefont
  {Guttridge}},\ and\ \bibinfo {author} {\bibfnamefont {S.~L.}\ \bibnamefont
  {Cornish}},\ }\href@noop {} {\bibfield  {journal} {\bibinfo  {journal}
  {Nature}\ }\textbf {\bibinfo {volume} {637}},\ \bibinfo {pages} {827–832}
  (\bibinfo {year} {2025})}\BibitemShut {NoStop}%
\bibitem [{\citenamefont {Picard}\ \emph {et~al.}(2025)\citenamefont {Picard},
  \citenamefont {Park}, \citenamefont {Patenotte}, \citenamefont
  {Gebretsadkan}, \citenamefont {Wellnitz}, \citenamefont {Rey},\ and\
  \citenamefont {Ni}}]{picard2025entanglement}%
  \BibitemOpen
  \bibfield  {author} {\bibinfo {author} {\bibfnamefont {L.~R.}\ \bibnamefont
  {Picard}}, \bibinfo {author} {\bibfnamefont {A.~J.}\ \bibnamefont {Park}},
  \bibinfo {author} {\bibfnamefont {G.~E.}\ \bibnamefont {Patenotte}}, \bibinfo
  {author} {\bibfnamefont {S.}~\bibnamefont {Gebretsadkan}}, \bibinfo {author}
  {\bibfnamefont {D.}~\bibnamefont {Wellnitz}}, \bibinfo {author}
  {\bibfnamefont {A.~M.}\ \bibnamefont {Rey}},\ and\ \bibinfo {author}
  {\bibfnamefont {K.-K.}\ \bibnamefont {Ni}},\ }\href@noop {} {\bibfield
  {journal} {\bibinfo  {journal} {Nature}\ }\textbf {\bibinfo {volume} {637}},\
  \bibinfo {pages} {821} (\bibinfo {year} {2025})}\BibitemShut {NoStop}%
\bibitem [{\citenamefont {Holland}\ \emph {et~al.}(2023)\citenamefont
  {Holland}, \citenamefont {Lu},\ and\ \citenamefont
  {Cheuk}}]{holland2023demand}%
  \BibitemOpen
  \bibfield  {author} {\bibinfo {author} {\bibfnamefont {C.~M.}\ \bibnamefont
  {Holland}}, \bibinfo {author} {\bibfnamefont {Y.}~\bibnamefont {Lu}},\ and\
  \bibinfo {author} {\bibfnamefont {L.~W.}\ \bibnamefont {Cheuk}},\ }\href@noop
  {} {\bibfield  {journal} {\bibinfo  {journal} {Science}\ }\textbf {\bibinfo
  {volume} {382}},\ \bibinfo {pages} {1143} (\bibinfo {year}
  {2023})}\BibitemShut {NoStop}%
\bibitem [{\citenamefont {Bao}\ \emph {et~al.}(2023)\citenamefont {Bao},
  \citenamefont {Yu}, \citenamefont {Anderegg}, \citenamefont {Chae},
  \citenamefont {Ketterle}, \citenamefont {Ni},\ and\ \citenamefont
  {Doyle}}]{bao2023dipolar}%
  \BibitemOpen
  \bibfield  {author} {\bibinfo {author} {\bibfnamefont {Y.}~\bibnamefont
  {Bao}}, \bibinfo {author} {\bibfnamefont {S.~S.}\ \bibnamefont {Yu}},
  \bibinfo {author} {\bibfnamefont {L.}~\bibnamefont {Anderegg}}, \bibinfo
  {author} {\bibfnamefont {E.}~\bibnamefont {Chae}}, \bibinfo {author}
  {\bibfnamefont {W.}~\bibnamefont {Ketterle}}, \bibinfo {author}
  {\bibfnamefont {K.-K.}\ \bibnamefont {Ni}},\ and\ \bibinfo {author}
  {\bibfnamefont {J.~M.}\ \bibnamefont {Doyle}},\ }\href@noop {} {\bibfield
  {journal} {\bibinfo  {journal} {Science}\ }\textbf {\bibinfo {volume}
  {382}},\ \bibinfo {pages} {1138} (\bibinfo {year} {2023})}\BibitemShut
  {NoStop}%
\bibitem [{\citenamefont {DeMille}\ \emph {et~al.}(2024)\citenamefont
  {DeMille}, \citenamefont {Hutzler}, \citenamefont {Rey},\ and\ \citenamefont
  {Zelevinsky}}]{demille2024quantum}%
  \BibitemOpen
  \bibfield  {author} {\bibinfo {author} {\bibfnamefont {D.}~\bibnamefont
  {DeMille}}, \bibinfo {author} {\bibfnamefont {N.~R.}\ \bibnamefont
  {Hutzler}}, \bibinfo {author} {\bibfnamefont {A.~M.}\ \bibnamefont {Rey}},\
  and\ \bibinfo {author} {\bibfnamefont {T.}~\bibnamefont {Zelevinsky}},\
  }\href@noop {} {\bibfield  {journal} {\bibinfo  {journal} {Nature Physics}\
  }\textbf {\bibinfo {volume} {20}},\ \bibinfo {pages} {741} (\bibinfo {year}
  {2024})}\BibitemShut {NoStop}%
\bibitem [{\citenamefont {Safronova}\ \emph {et~al.}(2018)\citenamefont
  {Safronova}, \citenamefont {Budker}, \citenamefont {DeMille}, \citenamefont
  {Kimball}, \citenamefont {Derevianko},\ and\ \citenamefont
  {Clark}}]{Safronova2018}%
  \BibitemOpen
  \bibfield  {author} {\bibinfo {author} {\bibfnamefont {M.~S.}\ \bibnamefont
  {Safronova}}, \bibinfo {author} {\bibfnamefont {D.}~\bibnamefont {Budker}},
  \bibinfo {author} {\bibfnamefont {D.}~\bibnamefont {DeMille}}, \bibinfo
  {author} {\bibfnamefont {D.~F.~J.}\ \bibnamefont {Kimball}}, \bibinfo
  {author} {\bibfnamefont {A.}~\bibnamefont {Derevianko}},\ and\ \bibinfo
  {author} {\bibfnamefont {C.~W.}\ \bibnamefont {Clark}},\ }\href
  {https://doi.org/10.1103/RevModPhys.90.025008} {\bibfield  {journal}
  {\bibinfo  {journal} {Rev. Mod. Phys.}\ }\textbf {\bibinfo {volume} {90}},\
  \bibinfo {pages} {025008} (\bibinfo {year} {2018})}\BibitemShut {NoStop}%
\bibitem [{\citenamefont {Langen}\ \emph {et~al.}(2024)\citenamefont {Langen},
  \citenamefont {Valtolina}, \citenamefont {Wang},\ and\ \citenamefont
  {Ye}}]{langen2024quantum}%
  \BibitemOpen
  \bibfield  {author} {\bibinfo {author} {\bibfnamefont {T.}~\bibnamefont
  {Langen}}, \bibinfo {author} {\bibfnamefont {G.}~\bibnamefont {Valtolina}},
  \bibinfo {author} {\bibfnamefont {D.}~\bibnamefont {Wang}},\ and\ \bibinfo
  {author} {\bibfnamefont {J.}~\bibnamefont {Ye}},\ }\href@noop {} {\bibfield
  {journal} {\bibinfo  {journal} {Nature Physics}\ }\textbf {\bibinfo {volume}
  {20}},\ \bibinfo {pages} {702} (\bibinfo {year} {2024})}\BibitemShut
  {NoStop}%
\bibitem [{\citenamefont {Wall}\ and\ \citenamefont
  {Carr}(2010)}]{wall2010hyperfine}%
  \BibitemOpen
  \bibfield  {author} {\bibinfo {author} {\bibfnamefont {M.~L.}\ \bibnamefont
  {Wall}}\ and\ \bibinfo {author} {\bibfnamefont {L.}~\bibnamefont {Carr}},\
  }\href@noop {} {\bibfield  {journal} {\bibinfo  {journal} {Phys. Rev. A}\
  }\textbf {\bibinfo {volume} {82}},\ \bibinfo {pages} {013611} (\bibinfo
  {year} {2010})}\BibitemShut {NoStop}%
\bibitem [{\citenamefont {Trefzger}\ \emph {et~al.}(2011)\citenamefont
  {Trefzger}, \citenamefont {Menotti}, \citenamefont {Capogrosso-Sansone},\
  and\ \citenamefont {Lewenstein}}]{trefzger2011ultracold}%
  \BibitemOpen
  \bibfield  {author} {\bibinfo {author} {\bibfnamefont {C.}~\bibnamefont
  {Trefzger}}, \bibinfo {author} {\bibfnamefont {C.}~\bibnamefont {Menotti}},
  \bibinfo {author} {\bibfnamefont {B.}~\bibnamefont {Capogrosso-Sansone}},\
  and\ \bibinfo {author} {\bibfnamefont {M.}~\bibnamefont {Lewenstein}},\
  }\href@noop {} {\bibfield  {journal} {\bibinfo  {journal} {J. Phys. B.}\
  }\textbf {\bibinfo {volume} {44}},\ \bibinfo {pages} {193001} (\bibinfo
  {year} {2011})}\BibitemShut {NoStop}%
\bibitem [{\citenamefont {Capogrosso-Sansone}\ \emph
  {et~al.}(2010)\citenamefont {Capogrosso-Sansone}, \citenamefont {Trefzger},
  \citenamefont {Lewenstein}, \citenamefont {Zoller},\ and\ \citenamefont
  {Pupillo}}]{capogrosso2010quantum}%
  \BibitemOpen
  \bibfield  {author} {\bibinfo {author} {\bibfnamefont {B.}~\bibnamefont
  {Capogrosso-Sansone}}, \bibinfo {author} {\bibfnamefont {C.}~\bibnamefont
  {Trefzger}}, \bibinfo {author} {\bibfnamefont {M.}~\bibnamefont
  {Lewenstein}}, \bibinfo {author} {\bibfnamefont {P.}~\bibnamefont {Zoller}},\
  and\ \bibinfo {author} {\bibfnamefont {G.}~\bibnamefont {Pupillo}},\
  }\href@noop {} {\bibfield  {journal} {\bibinfo  {journal} {Phys. Rev. Lett.}\
  }\textbf {\bibinfo {volume} {104}},\ \bibinfo {pages} {125301} (\bibinfo
  {year} {2010})}\BibitemShut {NoStop}%
\bibitem [{\citenamefont {Barnett}\ \emph {et~al.}(2006)\citenamefont
  {Barnett}, \citenamefont {Petrov}, \citenamefont {Lukin},\ and\ \citenamefont
  {Demler}}]{Barnett2006}%
  \BibitemOpen
  \bibfield  {author} {\bibinfo {author} {\bibfnamefont {R.}~\bibnamefont
  {Barnett}}, \bibinfo {author} {\bibfnamefont {D.}~\bibnamefont {Petrov}},
  \bibinfo {author} {\bibfnamefont {M.}~\bibnamefont {Lukin}},\ and\ \bibinfo
  {author} {\bibfnamefont {E.}~\bibnamefont {Demler}},\ }\href
  {https://doi.org/10.1103/PhysRevLett.96.190401} {\bibfield  {journal}
  {\bibinfo  {journal} {Phys. Rev. Lett.}\ }\textbf {\bibinfo {volume} {96}},\
  \bibinfo {pages} {190401} (\bibinfo {year} {2006})}\BibitemShut {NoStop}%
\bibitem [{\citenamefont {Carroll}\ \emph {et~al.}(2025)\citenamefont
  {Carroll}, \citenamefont {Hirzler}, \citenamefont {Miller}, \citenamefont
  {Wellnitz}, \citenamefont {Muleady}, \citenamefont {Lin}, \citenamefont
  {Zamarski}, \citenamefont {Wang}, \citenamefont {Bohn}, \citenamefont {Rey}
  \emph {et~al.}}]{carroll2025observation}%
  \BibitemOpen
  \bibfield  {author} {\bibinfo {author} {\bibfnamefont {A.~N.}\ \bibnamefont
  {Carroll}}, \bibinfo {author} {\bibfnamefont {H.}~\bibnamefont {Hirzler}},
  \bibinfo {author} {\bibfnamefont {C.}~\bibnamefont {Miller}}, \bibinfo
  {author} {\bibfnamefont {D.}~\bibnamefont {Wellnitz}}, \bibinfo {author}
  {\bibfnamefont {S.~R.}\ \bibnamefont {Muleady}}, \bibinfo {author}
  {\bibfnamefont {J.}~\bibnamefont {Lin}}, \bibinfo {author} {\bibfnamefont
  {K.~P.}\ \bibnamefont {Zamarski}}, \bibinfo {author} {\bibfnamefont {R.~R.}\
  \bibnamefont {Wang}}, \bibinfo {author} {\bibfnamefont {J.~L.}\ \bibnamefont
  {Bohn}}, \bibinfo {author} {\bibfnamefont {A.~M.}\ \bibnamefont {Rey}}, \emph
  {et~al.},\ }\href@noop {} {\bibfield  {journal} {\bibinfo  {journal}
  {Science}\ }\textbf {\bibinfo {volume} {388}},\ \bibinfo {pages} {381}
  (\bibinfo {year} {2025})}\BibitemShut {NoStop}%
\bibitem [{\citenamefont {Zhang}\ \emph {et~al.}(2025)\citenamefont {Zhang},
  \citenamefont {Liu}, \citenamefont {Deng}, \citenamefont {Chen},
  \citenamefont {Yi},\ and\ \citenamefont {Shi}}]{zhang2025supersolid}%
  \BibitemOpen
  \bibfield  {author} {\bibinfo {author} {\bibfnamefont {W.}~\bibnamefont
  {Zhang}}, \bibinfo {author} {\bibfnamefont {H.}~\bibnamefont {Liu}}, \bibinfo
  {author} {\bibfnamefont {F.}~\bibnamefont {Deng}}, \bibinfo {author}
  {\bibfnamefont {K.}~\bibnamefont {Chen}}, \bibinfo {author} {\bibfnamefont
  {S.}~\bibnamefont {Yi}},\ and\ \bibinfo {author} {\bibfnamefont
  {T.}~\bibnamefont {Shi}},\ }\href@noop {} {\bibfield  {journal} {\bibinfo
  {journal} {arXiv preprint arXiv:2506.23820}\ } (\bibinfo {year}
  {2025})}\BibitemShut {NoStop}%
\bibitem [{\citenamefont {Langen}\ \emph {et~al.}(2025)\citenamefont {Langen},
  \citenamefont {Boronat}, \citenamefont {S{\'a}nchez-Baena}, \citenamefont
  {Bomb{\'\i}n}, \citenamefont {Karman},\ and\ \citenamefont
  {Mazzanti}}]{langen2025dipolar}%
  \BibitemOpen
  \bibfield  {author} {\bibinfo {author} {\bibfnamefont {T.}~\bibnamefont
  {Langen}}, \bibinfo {author} {\bibfnamefont {J.}~\bibnamefont {Boronat}},
  \bibinfo {author} {\bibfnamefont {J.}~\bibnamefont {S{\'a}nchez-Baena}},
  \bibinfo {author} {\bibfnamefont {R.}~\bibnamefont {Bomb{\'\i}n}}, \bibinfo
  {author} {\bibfnamefont {T.}~\bibnamefont {Karman}},\ and\ \bibinfo {author}
  {\bibfnamefont {F.}~\bibnamefont {Mazzanti}},\ }\href@noop {} {\bibfield
  {journal} {\bibinfo  {journal} {Phys. Rev. Lett.}\ }\textbf {\bibinfo
  {volume} {134}},\ \bibinfo {pages} {053001} (\bibinfo {year}
  {2025})}\BibitemShut {NoStop}%
\bibitem [{\citenamefont {Ciardi}\ \emph {et~al.}(2025)\citenamefont {Ciardi},
  \citenamefont {Pedersen}, \citenamefont {Langen},\ and\ \citenamefont
  {Pohl}}]{ciardi2025self}%
  \BibitemOpen
  \bibfield  {author} {\bibinfo {author} {\bibfnamefont {M.}~\bibnamefont
  {Ciardi}}, \bibinfo {author} {\bibfnamefont {K.~R.}\ \bibnamefont
  {Pedersen}}, \bibinfo {author} {\bibfnamefont {T.}~\bibnamefont {Langen}},\
  and\ \bibinfo {author} {\bibfnamefont {T.}~\bibnamefont {Pohl}},\ }\href@noop
  {} {\bibfield  {journal} {\bibinfo  {journal} {Phys. Rev. Lett.}\ }\textbf
  {\bibinfo {volume} {135}},\ \bibinfo {pages} {153401} (\bibinfo {year}
  {2025})}\BibitemShut {NoStop}%
\bibitem [{\citenamefont {Mukherjee}\ \emph {et~al.}(2025)\citenamefont
  {Mukherjee}, \citenamefont {Hutson},\ and\ \citenamefont
  {Hazzard}}]{mukherjee2025n}%
  \BibitemOpen
  \bibfield  {author} {\bibinfo {author} {\bibfnamefont {B.}~\bibnamefont
  {Mukherjee}}, \bibinfo {author} {\bibfnamefont {J.~M.}\ \bibnamefont
  {Hutson}},\ and\ \bibinfo {author} {\bibfnamefont {K.~R.}\ \bibnamefont
  {Hazzard}},\ }\href@noop {} {\bibfield  {journal} {\bibinfo  {journal} {New.
  J. Phys.}\ }\textbf {\bibinfo {volume} {27}},\ \bibinfo {pages} {013013}
  (\bibinfo {year} {2025})}\BibitemShut {NoStop}%
\bibitem [{\citenamefont {Sundar}\ \emph {et~al.}(2018)\citenamefont {Sundar},
  \citenamefont {Gadway},\ and\ \citenamefont {Hazzard}}]{sundar2018synthetic}%
  \BibitemOpen
  \bibfield  {author} {\bibinfo {author} {\bibfnamefont {B.}~\bibnamefont
  {Sundar}}, \bibinfo {author} {\bibfnamefont {B.}~\bibnamefont {Gadway}},\
  and\ \bibinfo {author} {\bibfnamefont {K.~R.}\ \bibnamefont {Hazzard}},\
  }\href@noop {} {\bibfield  {journal} {\bibinfo  {journal} {Scientific
  reports}\ }\textbf {\bibinfo {volume} {8}},\ \bibinfo {pages} {3422}
  (\bibinfo {year} {2018})}\BibitemShut {NoStop}%
\bibitem [{\citenamefont {Feng}\ \emph {et~al.}(2022)\citenamefont {Feng},
  \citenamefont {Manetsch}, \citenamefont {Rousseau}, \citenamefont {Hazzard},\
  and\ \citenamefont {Scalettar}}]{feng2022quantum}%
  \BibitemOpen
  \bibfield  {author} {\bibinfo {author} {\bibfnamefont {C.}~\bibnamefont
  {Feng}}, \bibinfo {author} {\bibfnamefont {H.}~\bibnamefont {Manetsch}},
  \bibinfo {author} {\bibfnamefont {V.~G.}\ \bibnamefont {Rousseau}}, \bibinfo
  {author} {\bibfnamefont {K.~R.}\ \bibnamefont {Hazzard}},\ and\ \bibinfo
  {author} {\bibfnamefont {R.}~\bibnamefont {Scalettar}},\ }\href@noop {}
  {\bibfield  {journal} {\bibinfo  {journal} {Phys. Rev. A}\ }\textbf {\bibinfo
  {volume} {105}},\ \bibinfo {pages} {063320} (\bibinfo {year}
  {2022})}\BibitemShut {NoStop}%
\bibitem [{\citenamefont {De~Marco}\ \emph {et~al.}(2019)\citenamefont
  {De~Marco}, \citenamefont {Valtolina}, \citenamefont {Matsuda}, \citenamefont
  {Tobias}, \citenamefont {Covey},\ and\ \citenamefont {Ye}}]{demarco:19}%
  \BibitemOpen
  \bibfield  {author} {\bibinfo {author} {\bibfnamefont {L.}~\bibnamefont
  {De~Marco}}, \bibinfo {author} {\bibfnamefont {G.}~\bibnamefont {Valtolina}},
  \bibinfo {author} {\bibfnamefont {K.}~\bibnamefont {Matsuda}}, \bibinfo
  {author} {\bibfnamefont {W.~G.}\ \bibnamefont {Tobias}}, \bibinfo {author}
  {\bibfnamefont {J.~P.}\ \bibnamefont {Covey}},\ and\ \bibinfo {author}
  {\bibfnamefont {J.}~\bibnamefont {Ye}},\ }\href
  {https://doi.org/10.1126/science.aau7230} {\bibfield  {journal} {\bibinfo
  {journal} {Science}\ }\textbf {\bibinfo {volume} {363}},\ \bibinfo {pages}
  {853} (\bibinfo {year} {2019})}\BibitemShut {NoStop}%
\bibitem [{\citenamefont {Valtolina}\ \emph {et~al.}(2020)\citenamefont
  {Valtolina}, \citenamefont {Matsuda}, \citenamefont {Tobias}, \citenamefont
  {Li}, \citenamefont {De~Marco},\ and\ \citenamefont {Ye}}]{valtolina:20}%
  \BibitemOpen
  \bibfield  {author} {\bibinfo {author} {\bibfnamefont {G.}~\bibnamefont
  {Valtolina}}, \bibinfo {author} {\bibfnamefont {K.}~\bibnamefont {Matsuda}},
  \bibinfo {author} {\bibfnamefont {W.~G.}\ \bibnamefont {Tobias}}, \bibinfo
  {author} {\bibfnamefont {J.-R.}\ \bibnamefont {Li}}, \bibinfo {author}
  {\bibfnamefont {L.}~\bibnamefont {De~Marco}},\ and\ \bibinfo {author}
  {\bibfnamefont {J.}~\bibnamefont {Ye}},\ }\href
  {https://doi.org/10.1038/s41586-020-2980-7} {\bibfield  {journal} {\bibinfo
  {journal} {Nature}\ }\textbf {\bibinfo {volume} {588}},\ \bibinfo {pages}
  {239} (\bibinfo {year} {2020})}\BibitemShut {NoStop}%
\bibitem [{\citenamefont {Matsuda}\ \emph {et~al.}(2020)\citenamefont
  {Matsuda}, \citenamefont {De~Marco}, \citenamefont {Li}, \citenamefont
  {Tobias}, \citenamefont {Valtolina}, \citenamefont {Qu{\'e}m{\'e}ner},\ and\
  \citenamefont {Ye}}]{matsuda:20}%
  \BibitemOpen
  \bibfield  {author} {\bibinfo {author} {\bibfnamefont {K.}~\bibnamefont
  {Matsuda}}, \bibinfo {author} {\bibfnamefont {L.}~\bibnamefont {De~Marco}},
  \bibinfo {author} {\bibfnamefont {J.-R.}\ \bibnamefont {Li}}, \bibinfo
  {author} {\bibfnamefont {W.~G.}\ \bibnamefont {Tobias}}, \bibinfo {author}
  {\bibfnamefont {G.}~\bibnamefont {Valtolina}}, \bibinfo {author}
  {\bibfnamefont {G.}~\bibnamefont {Qu{\'e}m{\'e}ner}},\ and\ \bibinfo {author}
  {\bibfnamefont {J.}~\bibnamefont {Ye}},\ }\href
  {https://doi.org/10.1126/science.abe7370} {\bibfield  {journal} {\bibinfo
  {journal} {Science}\ }\textbf {\bibinfo {volume} {370}},\ \bibinfo {pages}
  {1324} (\bibinfo {year} {2020})}\BibitemShut {NoStop}%
\bibitem [{\citenamefont {Schindewolf}\ \emph {et~al.}(2022)\citenamefont
  {Schindewolf}, \citenamefont {Bause}, \citenamefont {Chen}, \citenamefont
  {Duda}, \citenamefont {Karman}, \citenamefont {Bloch},\ and\ \citenamefont
  {Luo}}]{schindewolf:22}%
  \BibitemOpen
  \bibfield  {author} {\bibinfo {author} {\bibfnamefont {A.}~\bibnamefont
  {Schindewolf}}, \bibinfo {author} {\bibfnamefont {R.}~\bibnamefont {Bause}},
  \bibinfo {author} {\bibfnamefont {X.-Y.}\ \bibnamefont {Chen}}, \bibinfo
  {author} {\bibfnamefont {M.}~\bibnamefont {Duda}}, \bibinfo {author}
  {\bibfnamefont {T.}~\bibnamefont {Karman}}, \bibinfo {author} {\bibfnamefont
  {I.}~\bibnamefont {Bloch}},\ and\ \bibinfo {author} {\bibfnamefont {X.-Y.}\
  \bibnamefont {Luo}},\ }\href {https://doi.org/10.1038/s41586-022-04900-0}
  {\bibfield  {journal} {\bibinfo  {journal} {Nature}\ }\textbf {\bibinfo
  {volume} {607}},\ \bibinfo {pages} {677} (\bibinfo {year}
  {2022})}\BibitemShut {NoStop}%
\bibitem [{\citenamefont {Bigagli}\ \emph {et~al.}(2024)\citenamefont
  {Bigagli}, \citenamefont {Yuan}, \citenamefont {Zhang}, \citenamefont
  {Bulatovic}, \citenamefont {Karman}, \citenamefont {Stevenson},\ and\
  \citenamefont {Will}}]{bigagli:24}%
  \BibitemOpen
  \bibfield  {author} {\bibinfo {author} {\bibfnamefont {N.}~\bibnamefont
  {Bigagli}}, \bibinfo {author} {\bibfnamefont {W.}~\bibnamefont {Yuan}},
  \bibinfo {author} {\bibfnamefont {S.}~\bibnamefont {Zhang}}, \bibinfo
  {author} {\bibfnamefont {B.}~\bibnamefont {Bulatovic}}, \bibinfo {author}
  {\bibfnamefont {T.}~\bibnamefont {Karman}}, \bibinfo {author} {\bibfnamefont
  {I.}~\bibnamefont {Stevenson}},\ and\ \bibinfo {author} {\bibfnamefont
  {S.}~\bibnamefont {Will}},\ }\href
  {https://doi.org/10.1038/s41586-024-07492-z} {\bibfield  {journal} {\bibinfo
  {journal} {Nature}\ }\textbf {\bibinfo {volume} {631}},\ \bibinfo {pages}
  {289} (\bibinfo {year} {2024})}\BibitemShut {NoStop}%
\bibitem [{\citenamefont {Shi}\ \emph {et~al.}(2025)\citenamefont {Shi},
  \citenamefont {Huang}, \citenamefont {Deng}, \citenamefont {Jin},
  \citenamefont {Yi}, \citenamefont {Shi},\ and\ \citenamefont
  {Wang}}]{shi2025boseeinsteincondensateultracoldsodiumrubidium}%
  \BibitemOpen
  \bibfield  {author} {\bibinfo {author} {\bibfnamefont {Z.}~\bibnamefont
  {Shi}}, \bibinfo {author} {\bibfnamefont {Z.}~\bibnamefont {Huang}}, \bibinfo
  {author} {\bibfnamefont {F.}~\bibnamefont {Deng}}, \bibinfo {author}
  {\bibfnamefont {W.-J.}\ \bibnamefont {Jin}}, \bibinfo {author} {\bibfnamefont
  {S.}~\bibnamefont {Yi}}, \bibinfo {author} {\bibfnamefont {T.}~\bibnamefont
  {Shi}},\ and\ \bibinfo {author} {\bibfnamefont {D.}~\bibnamefont {Wang}},\
  }\href {https://arxiv.org/abs/2508.20518} {\bibinfo {title} {Bose-einstein
  condensate of ultracold sodium-rubidium molecules with tunable dipolar
  interactions}} (\bibinfo {year} {2025}),\ \Eprint
  {https://arxiv.org/abs/2508.20518} {arXiv:2508.20518 [cond-mat.quant-gas]}
  \BibitemShut {NoStop}%
\bibitem [{\citenamefont {Yuan}\ \emph {et~al.}(2025)\citenamefont {Yuan},
  \citenamefont {Zhang}, \citenamefont {Bigagli}, \citenamefont {Kwak},
  \citenamefont {Warner}, \citenamefont {Karman}, \citenamefont {Stevenson},\
  and\ \citenamefont {Will}}]{yuan:26}%
  \BibitemOpen
  \bibfield  {author} {\bibinfo {author} {\bibfnamefont {W.}~\bibnamefont
  {Yuan}}, \bibinfo {author} {\bibfnamefont {S.}~\bibnamefont {Zhang}},
  \bibinfo {author} {\bibfnamefont {N.}~\bibnamefont {Bigagli}}, \bibinfo
  {author} {\bibfnamefont {H.}~\bibnamefont {Kwak}}, \bibinfo {author}
  {\bibfnamefont {C.}~\bibnamefont {Warner}}, \bibinfo {author} {\bibfnamefont
  {T.}~\bibnamefont {Karman}}, \bibinfo {author} {\bibfnamefont
  {I.}~\bibnamefont {Stevenson}},\ and\ \bibinfo {author} {\bibfnamefont
  {S.}~\bibnamefont {Will}},\ }\href@noop {} {\bibfield  {journal} {\bibinfo
  {journal} {arXiv preprint arXiv:2505.08773}\ } (\bibinfo {year}
  {2025})}\BibitemShut {NoStop}%
\bibitem [{\citenamefont {Karman}\ \emph {et~al.}(2025)\citenamefont {Karman},
  \citenamefont {Bigagli}, \citenamefont {Yuan}, \citenamefont {Zhang},
  \citenamefont {Stevenson},\ and\ \citenamefont {Will}}]{karman:25}%
  \BibitemOpen
  \bibfield  {author} {\bibinfo {author} {\bibfnamefont {T.}~\bibnamefont
  {Karman}}, \bibinfo {author} {\bibfnamefont {N.}~\bibnamefont {Bigagli}},
  \bibinfo {author} {\bibfnamefont {W.}~\bibnamefont {Yuan}}, \bibinfo {author}
  {\bibfnamefont {S.}~\bibnamefont {Zhang}}, \bibinfo {author} {\bibfnamefont
  {I.}~\bibnamefont {Stevenson}},\ and\ \bibinfo {author} {\bibfnamefont
  {S.}~\bibnamefont {Will}},\ }\href {https://doi.org/10.1103/b8pm-3prn}
  {\bibfield  {journal} {\bibinfo  {journal} {PRX Quantum}\ }\textbf {\bibinfo
  {volume} {6}},\ \bibinfo {pages} {020358} (\bibinfo {year}
  {2025})}\BibitemShut {NoStop}%
\bibitem [{\citenamefont {Chen}\ \emph {et~al.}(2023)\citenamefont {Chen},
  \citenamefont {Schindewolf}, \citenamefont {Eppelt}, \citenamefont {Bause},
  \citenamefont {Duda}, \citenamefont {Biswas}, \citenamefont {Karman},
  \citenamefont {Hilker}, \citenamefont {Bloch},\ and\ \citenamefont
  {Luo}}]{chen:23}%
  \BibitemOpen
  \bibfield  {author} {\bibinfo {author} {\bibfnamefont {X.-Y.}\ \bibnamefont
  {Chen}}, \bibinfo {author} {\bibfnamefont {A.}~\bibnamefont {Schindewolf}},
  \bibinfo {author} {\bibfnamefont {S.}~\bibnamefont {Eppelt}}, \bibinfo
  {author} {\bibfnamefont {R.}~\bibnamefont {Bause}}, \bibinfo {author}
  {\bibfnamefont {M.}~\bibnamefont {Duda}}, \bibinfo {author} {\bibfnamefont
  {S.}~\bibnamefont {Biswas}}, \bibinfo {author} {\bibfnamefont
  {T.}~\bibnamefont {Karman}}, \bibinfo {author} {\bibfnamefont
  {T.}~\bibnamefont {Hilker}}, \bibinfo {author} {\bibfnamefont
  {I.}~\bibnamefont {Bloch}},\ and\ \bibinfo {author} {\bibfnamefont {X.-Y.}\
  \bibnamefont {Luo}},\ }\href {https://doi.org/10.1038/s41586-022-05651-8}
  {\bibfield  {journal} {\bibinfo  {journal} {Nature}\ }\textbf {\bibinfo
  {volume} {614}},\ \bibinfo {pages} {59} (\bibinfo {year} {2023})}\BibitemShut
  {NoStop}%
\bibitem [{\citenamefont {Zhang}\ \emph {et~al.}(2026)\citenamefont {Zhang},
  \citenamefont {Yuan}, \citenamefont {Bigagli}, \citenamefont {Kwak},
  \citenamefont {Karman}, \citenamefont {Stevenson},\ and\ \citenamefont
  {Will}}]{zhang:26}%
  \BibitemOpen
  \bibfield  {author} {\bibinfo {author} {\bibfnamefont {S.}~\bibnamefont
  {Zhang}}, \bibinfo {author} {\bibfnamefont {W.}~\bibnamefont {Yuan}},
  \bibinfo {author} {\bibfnamefont {N.}~\bibnamefont {Bigagli}}, \bibinfo
  {author} {\bibfnamefont {H.}~\bibnamefont {Kwak}}, \bibinfo {author}
  {\bibfnamefont {T.}~\bibnamefont {Karman}}, \bibinfo {author} {\bibfnamefont
  {I.}~\bibnamefont {Stevenson}},\ and\ \bibinfo {author} {\bibfnamefont
  {S.}~\bibnamefont {Will}},\ }\href
  {https://doi.org/10.1038/s41586-026-10245-9} {\bibfield  {journal} {\bibinfo
  {journal} {Nature}\ }\textbf {\bibinfo {volume} {651}},\ \bibinfo {pages}
  {601–606} (\bibinfo {year} {2026})}\BibitemShut {NoStop}%
\bibitem [{\citenamefont {Schindewolf}\ \emph {et~al.}(2025)\citenamefont
  {Schindewolf}, \citenamefont {Hertkorn}, \citenamefont {Stevenson},
  \citenamefont {Ciardi}, \citenamefont {Gross}, \citenamefont {Wang},
  \citenamefont {Karman}, \citenamefont {Quemener}, \citenamefont {Will},
  \citenamefont {Pohl} \emph {et~al.}}]{schindewolf2025few}%
  \BibitemOpen
  \bibfield  {author} {\bibinfo {author} {\bibfnamefont {A.}~\bibnamefont
  {Schindewolf}}, \bibinfo {author} {\bibfnamefont {J.}~\bibnamefont
  {Hertkorn}}, \bibinfo {author} {\bibfnamefont {I.}~\bibnamefont {Stevenson}},
  \bibinfo {author} {\bibfnamefont {M.}~\bibnamefont {Ciardi}}, \bibinfo
  {author} {\bibfnamefont {P.}~\bibnamefont {Gross}}, \bibinfo {author}
  {\bibfnamefont {D.}~\bibnamefont {Wang}}, \bibinfo {author} {\bibfnamefont
  {T.}~\bibnamefont {Karman}}, \bibinfo {author} {\bibfnamefont
  {G.}~\bibnamefont {Quemener}}, \bibinfo {author} {\bibfnamefont
  {S.}~\bibnamefont {Will}}, \bibinfo {author} {\bibfnamefont {T.}~\bibnamefont
  {Pohl}}, \emph {et~al.},\ }\href@noop {} {\bibfield  {journal} {\bibinfo
  {journal} {arXiv preprint arXiv:2512.14511}\ } (\bibinfo {year}
  {2025})}\BibitemShut {NoStop}%
\bibitem [{\citenamefont {Walraven}\ and\ \citenamefont
  {Karman}(2024)}]{walraven:24a}%
  \BibitemOpen
  \bibfield  {author} {\bibinfo {author} {\bibfnamefont {E.~F.}\ \bibnamefont
  {Walraven}}\ and\ \bibinfo {author} {\bibfnamefont {T.}~\bibnamefont
  {Karman}},\ }\href {https://doi.org/10.1103/PhysRevA.109.043310} {\bibfield
  {journal} {\bibinfo  {journal} {Phys. Rev. A}\ }\textbf {\bibinfo {volume}
  {109}},\ \bibinfo {pages} {043310} (\bibinfo {year} {2024})}\BibitemShut
  {NoStop}%
\bibitem [{\citenamefont {Walraven}\ and\ \citenamefont
  {Karman}(2025)}]{walraven:25}%
  \BibitemOpen
  \bibfield  {author} {\bibinfo {author} {\bibfnamefont {E.~F.}\ \bibnamefont
  {Walraven}}\ and\ \bibinfo {author} {\bibfnamefont {T.}~\bibnamefont
  {Karman}},\ }\href {https://doi.org/10.1103/z2bv-9gw7} {\bibfield  {journal}
  {\bibinfo  {journal} {Phys. Rev. A}\ }\textbf {\bibinfo {volume} {112}},\
  \bibinfo {pages} {032810} (\bibinfo {year} {2025})}\BibitemShut {NoStop}%
\bibitem [{\citenamefont {Ye}\ \emph {et~al.}(2018)\citenamefont {Ye},
  \citenamefont {Guo}, \citenamefont {Gonz{\'a}lez-Mart{\'\i}nez},
  \citenamefont {Qu{\'e}m{\'e}ner},\ and\ \citenamefont
  {Wang}}]{ye2018collisions}%
  \BibitemOpen
  \bibfield  {author} {\bibinfo {author} {\bibfnamefont {X.}~\bibnamefont
  {Ye}}, \bibinfo {author} {\bibfnamefont {M.}~\bibnamefont {Guo}}, \bibinfo
  {author} {\bibfnamefont {M.~L.}\ \bibnamefont {Gonz{\'a}lez-Mart{\'\i}nez}},
  \bibinfo {author} {\bibfnamefont {G.}~\bibnamefont {Qu{\'e}m{\'e}ner}},\ and\
  \bibinfo {author} {\bibfnamefont {D.}~\bibnamefont {Wang}},\ }\href@noop {}
  {\bibfield  {journal} {\bibinfo  {journal} {Science advances}\ }\textbf
  {\bibinfo {volume} {4}},\ \bibinfo {pages} {eaaq0083} (\bibinfo {year}
  {2018})}\BibitemShut {NoStop}%
\bibitem [{\citenamefont {Kozyryev}\ and\ \citenamefont
  {Hutzler}(2017)}]{kozyryev2017precision}%
  \BibitemOpen
  \bibfield  {author} {\bibinfo {author} {\bibfnamefont {I.}~\bibnamefont
  {Kozyryev}}\ and\ \bibinfo {author} {\bibfnamefont {N.~R.}\ \bibnamefont
  {Hutzler}},\ }\href@noop {} {\bibfield  {journal} {\bibinfo  {journal} {Phys.
  Rev. Lett.}\ }\textbf {\bibinfo {volume} {119}},\ \bibinfo {pages} {133002}
  (\bibinfo {year} {2017})}\BibitemShut {NoStop}%
\bibitem [{\citenamefont {Hutzler}(2020)}]{hutzler2020polyatomic}%
  \BibitemOpen
  \bibfield  {author} {\bibinfo {author} {\bibfnamefont {N.~R.}\ \bibnamefont
  {Hutzler}},\ }\href@noop {} {\bibfield  {journal} {\bibinfo  {journal}
  {Quantum Science \& Technology}\ }\textbf {\bibinfo {volume} {5}},\ \bibinfo
  {pages} {044011} (\bibinfo {year} {2020})}\BibitemShut {NoStop}%
\bibitem [{\citenamefont {Anderegg}\ \emph {et~al.}(2023)\citenamefont
  {Anderegg}, \citenamefont {Vilas}, \citenamefont {Hallas}, \citenamefont
  {Robichaud}, \citenamefont {Jadbabaie}, \citenamefont {Doyle},\ and\
  \citenamefont {Hutzler}}]{anderegg2023quantum}%
  \BibitemOpen
  \bibfield  {author} {\bibinfo {author} {\bibfnamefont {L.}~\bibnamefont
  {Anderegg}}, \bibinfo {author} {\bibfnamefont {N.~B.}\ \bibnamefont {Vilas}},
  \bibinfo {author} {\bibfnamefont {C.}~\bibnamefont {Hallas}}, \bibinfo
  {author} {\bibfnamefont {P.}~\bibnamefont {Robichaud}}, \bibinfo {author}
  {\bibfnamefont {A.}~\bibnamefont {Jadbabaie}}, \bibinfo {author}
  {\bibfnamefont {J.~M.}\ \bibnamefont {Doyle}},\ and\ \bibinfo {author}
  {\bibfnamefont {N.~R.}\ \bibnamefont {Hutzler}},\ }\href@noop {} {\bibfield
  {journal} {\bibinfo  {journal} {Science}\ }\textbf {\bibinfo {volume}
  {382}},\ \bibinfo {pages} {665} (\bibinfo {year} {2023})}\BibitemShut
  {NoStop}%
\bibitem [{\citenamefont {Augustovi{\v{c}}ov{\'a}}\ and\ \citenamefont
  {Bohn}(2019)}]{augustovivcova2019ultracold}%
  \BibitemOpen
  \bibfield  {author} {\bibinfo {author} {\bibfnamefont {L.~D.}\ \bibnamefont
  {Augustovi{\v{c}}ov{\'a}}}\ and\ \bibinfo {author} {\bibfnamefont {J.~L.}\
  \bibnamefont {Bohn}},\ }\href@noop {} {\bibfield  {journal} {\bibinfo
  {journal} {New. J. Phys.}\ }\textbf {\bibinfo {volume} {21}},\ \bibinfo
  {pages} {103022} (\bibinfo {year} {2019})}\BibitemShut {NoStop}%
\bibitem [{\citenamefont {Vilas}\ \emph {et~al.}(2026)\citenamefont {Vilas},
  \citenamefont {Robichaud}, \citenamefont {Hallas}, \citenamefont {Tao},
  \citenamefont {Anderegg}, \citenamefont {Li}, \citenamefont {Lampson},
  \citenamefont {Augustovi{\v{c}}ov{\'a}}, \citenamefont {Bohn},\ and\
  \citenamefont {Doyle}}]{vilas2026quantum}%
  \BibitemOpen
  \bibfield  {author} {\bibinfo {author} {\bibfnamefont {N.~B.}\ \bibnamefont
  {Vilas}}, \bibinfo {author} {\bibfnamefont {P.}~\bibnamefont {Robichaud}},
  \bibinfo {author} {\bibfnamefont {C.}~\bibnamefont {Hallas}}, \bibinfo
  {author} {\bibfnamefont {J.}~\bibnamefont {Tao}}, \bibinfo {author}
  {\bibfnamefont {L.}~\bibnamefont {Anderegg}}, \bibinfo {author}
  {\bibfnamefont {G.~K.}\ \bibnamefont {Li}}, \bibinfo {author} {\bibfnamefont
  {H.}~\bibnamefont {Lampson}}, \bibinfo {author} {\bibfnamefont {L.~D.}\
  \bibnamefont {Augustovi{\v{c}}ov{\'a}}}, \bibinfo {author} {\bibfnamefont
  {J.~L.}\ \bibnamefont {Bohn}},\ and\ \bibinfo {author} {\bibfnamefont
  {J.~M.}\ \bibnamefont {Doyle}},\ }\href@noop {} {\bibfield  {journal}
  {\bibinfo  {journal} {Phys. Rev. X}\ }\textbf {\bibinfo {volume} {16}},\
  \bibinfo {pages} {021001} (\bibinfo {year} {2026})}\BibitemShut {NoStop}%
\bibitem [{\citenamefont {Ni}\ \emph {et~al.}(2008)\citenamefont {Ni},
  \citenamefont {Ospelkaus}, \citenamefont {De~Miranda}, \citenamefont {Pe'Er},
  \citenamefont {Neyenhuis}, \citenamefont {Zirbel}, \citenamefont
  {Kotochigova}, \citenamefont {Julienne}, \citenamefont {Jin},\ and\
  \citenamefont {Ye}}]{ni2008high}%
  \BibitemOpen
  \bibfield  {author} {\bibinfo {author} {\bibfnamefont {K.-K.}\ \bibnamefont
  {Ni}}, \bibinfo {author} {\bibfnamefont {S.}~\bibnamefont {Ospelkaus}},
  \bibinfo {author} {\bibfnamefont {M.}~\bibnamefont {De~Miranda}}, \bibinfo
  {author} {\bibfnamefont {A.}~\bibnamefont {Pe'Er}}, \bibinfo {author}
  {\bibfnamefont {B.}~\bibnamefont {Neyenhuis}}, \bibinfo {author}
  {\bibfnamefont {J.}~\bibnamefont {Zirbel}}, \bibinfo {author} {\bibfnamefont
  {S.}~\bibnamefont {Kotochigova}}, \bibinfo {author} {\bibfnamefont
  {P.}~\bibnamefont {Julienne}}, \bibinfo {author} {\bibfnamefont
  {D.}~\bibnamefont {Jin}},\ and\ \bibinfo {author} {\bibfnamefont
  {J.}~\bibnamefont {Ye}},\ }\href@noop {} {\bibfield  {journal} {\bibinfo
  {journal} {Science}\ }\textbf {\bibinfo {volume} {322}},\ \bibinfo {pages}
  {231} (\bibinfo {year} {2008})}\BibitemShut {NoStop}%
\bibitem [{\citenamefont {Takekoshi}\ \emph {et~al.}(2014)\citenamefont
  {Takekoshi}, \citenamefont {Reichs{\"o}llner}, \citenamefont {Schindewolf},
  \citenamefont {Hutson}, \citenamefont {Le~Sueur}, \citenamefont {Dulieu},
  \citenamefont {Ferlaino}, \citenamefont {Grimm},\ and\ \citenamefont
  {N{\"a}gerl}}]{takekoshi2014ultracold}%
  \BibitemOpen
  \bibfield  {author} {\bibinfo {author} {\bibfnamefont {T.}~\bibnamefont
  {Takekoshi}}, \bibinfo {author} {\bibfnamefont {L.}~\bibnamefont
  {Reichs{\"o}llner}}, \bibinfo {author} {\bibfnamefont {A.}~\bibnamefont
  {Schindewolf}}, \bibinfo {author} {\bibfnamefont {J.~M.}\ \bibnamefont
  {Hutson}}, \bibinfo {author} {\bibfnamefont {C.~R.}\ \bibnamefont
  {Le~Sueur}}, \bibinfo {author} {\bibfnamefont {O.}~\bibnamefont {Dulieu}},
  \bibinfo {author} {\bibfnamefont {F.}~\bibnamefont {Ferlaino}}, \bibinfo
  {author} {\bibfnamefont {R.}~\bibnamefont {Grimm}},\ and\ \bibinfo {author}
  {\bibfnamefont {H.-C.}\ \bibnamefont {N{\"a}gerl}},\ }\href@noop {}
  {\bibfield  {journal} {\bibinfo  {journal} {Phys. Rev. Lett.}\ }\textbf
  {\bibinfo {volume} {113}},\ \bibinfo {pages} {205301} (\bibinfo {year}
  {2014})}\BibitemShut {NoStop}%
\bibitem [{\citenamefont {Molony}\ \emph {et~al.}(2014)\citenamefont {Molony},
  \citenamefont {Gregory}, \citenamefont {Ji}, \citenamefont {Lu},
  \citenamefont {K{\"o}ppinger}, \citenamefont {Le~Sueur}, \citenamefont
  {Blackley}, \citenamefont {Hutson},\ and\ \citenamefont
  {Cornish}}]{molony2014creation}%
  \BibitemOpen
  \bibfield  {author} {\bibinfo {author} {\bibfnamefont {P.~K.}\ \bibnamefont
  {Molony}}, \bibinfo {author} {\bibfnamefont {P.~D.}\ \bibnamefont {Gregory}},
  \bibinfo {author} {\bibfnamefont {Z.}~\bibnamefont {Ji}}, \bibinfo {author}
  {\bibfnamefont {B.}~\bibnamefont {Lu}}, \bibinfo {author} {\bibfnamefont
  {M.~P.}\ \bibnamefont {K{\"o}ppinger}}, \bibinfo {author} {\bibfnamefont
  {C.~R.}\ \bibnamefont {Le~Sueur}}, \bibinfo {author} {\bibfnamefont {C.~L.}\
  \bibnamefont {Blackley}}, \bibinfo {author} {\bibfnamefont {J.~M.}\
  \bibnamefont {Hutson}},\ and\ \bibinfo {author} {\bibfnamefont {S.~L.}\
  \bibnamefont {Cornish}},\ }\href@noop {} {\bibfield  {journal} {\bibinfo
  {journal} {Phys. Rev. Lett.}\ }\textbf {\bibinfo {volume} {113}},\ \bibinfo
  {pages} {255301} (\bibinfo {year} {2014})}\BibitemShut {NoStop}%
\bibitem [{\citenamefont {Park}\ \emph {et~al.}(2015)\citenamefont {Park},
  \citenamefont {Will},\ and\ \citenamefont {Zwierlein}}]{park2015ultracold}%
  \BibitemOpen
  \bibfield  {author} {\bibinfo {author} {\bibfnamefont {J.~W.}\ \bibnamefont
  {Park}}, \bibinfo {author} {\bibfnamefont {S.~A.}\ \bibnamefont {Will}},\
  and\ \bibinfo {author} {\bibfnamefont {M.~W.}\ \bibnamefont {Zwierlein}},\
  }\href@noop {} {\bibfield  {journal} {\bibinfo  {journal} {Phys. Rev. Lett.}\
  }\textbf {\bibinfo {volume} {114}},\ \bibinfo {pages} {205302} (\bibinfo
  {year} {2015})}\BibitemShut {NoStop}%
\bibitem [{\citenamefont {Guo}\ \emph {et~al.}(2016)\citenamefont {Guo},
  \citenamefont {Zhu}, \citenamefont {Lu}, \citenamefont {Ye}, \citenamefont
  {Wang}, \citenamefont {Vexiau}, \citenamefont {Bouloufa-Maafa}, \citenamefont
  {Qu{\'e}m{\'e}ner}, \citenamefont {Dulieu},\ and\ \citenamefont
  {Wang}}]{guo2016creation}%
  \BibitemOpen
  \bibfield  {author} {\bibinfo {author} {\bibfnamefont {M.}~\bibnamefont
  {Guo}}, \bibinfo {author} {\bibfnamefont {B.}~\bibnamefont {Zhu}}, \bibinfo
  {author} {\bibfnamefont {B.}~\bibnamefont {Lu}}, \bibinfo {author}
  {\bibfnamefont {X.}~\bibnamefont {Ye}}, \bibinfo {author} {\bibfnamefont
  {F.}~\bibnamefont {Wang}}, \bibinfo {author} {\bibfnamefont {R.}~\bibnamefont
  {Vexiau}}, \bibinfo {author} {\bibfnamefont {N.}~\bibnamefont
  {Bouloufa-Maafa}}, \bibinfo {author} {\bibfnamefont {G.}~\bibnamefont
  {Qu{\'e}m{\'e}ner}}, \bibinfo {author} {\bibfnamefont {O.}~\bibnamefont
  {Dulieu}},\ and\ \bibinfo {author} {\bibfnamefont {D.}~\bibnamefont {Wang}},\
  }\href@noop {} {\bibfield  {journal} {\bibinfo  {journal} {Phys. Rev. Lett.}\
  }\textbf {\bibinfo {volume} {116}},\ \bibinfo {pages} {205303} (\bibinfo
  {year} {2016})}\BibitemShut {NoStop}%
\bibitem [{\citenamefont {Stevenson}\ \emph {et~al.}(2023)\citenamefont
  {Stevenson}, \citenamefont {Lam}, \citenamefont {Bigagli}, \citenamefont
  {Warner}, \citenamefont {Yuan}, \citenamefont {Zhang},\ and\ \citenamefont
  {Will}}]{Stevenson2023}%
  \BibitemOpen
  \bibfield  {author} {\bibinfo {author} {\bibfnamefont {I.}~\bibnamefont
  {Stevenson}}, \bibinfo {author} {\bibfnamefont {A.~Z.}\ \bibnamefont {Lam}},
  \bibinfo {author} {\bibfnamefont {N.}~\bibnamefont {Bigagli}}, \bibinfo
  {author} {\bibfnamefont {C.}~\bibnamefont {Warner}}, \bibinfo {author}
  {\bibfnamefont {W.}~\bibnamefont {Yuan}}, \bibinfo {author} {\bibfnamefont
  {S.}~\bibnamefont {Zhang}},\ and\ \bibinfo {author} {\bibfnamefont
  {S.}~\bibnamefont {Will}},\ }\href
  {https://doi.org/10.1103/PhysRevLett.130.113002} {\bibfield  {journal}
  {\bibinfo  {journal} {Phys. Rev. Lett.}\ }\textbf {\bibinfo {volume} {130}},\
  \bibinfo {pages} {113002} (\bibinfo {year} {2023})}\BibitemShut {NoStop}%
\bibitem [{\citenamefont {Zhelyazkova}\ \emph {et~al.}(2014)\citenamefont
  {Zhelyazkova}, \citenamefont {Cournol}, \citenamefont {Wall}, \citenamefont
  {Matsushima}, \citenamefont {Hudson}, \citenamefont {Hinds}, \citenamefont
  {Tarbutt},\ and\ \citenamefont {Sauer}}]{zhelyazkova2014laser}%
  \BibitemOpen
  \bibfield  {author} {\bibinfo {author} {\bibfnamefont {V.}~\bibnamefont
  {Zhelyazkova}}, \bibinfo {author} {\bibfnamefont {A.}~\bibnamefont
  {Cournol}}, \bibinfo {author} {\bibfnamefont {T.~E.}\ \bibnamefont {Wall}},
  \bibinfo {author} {\bibfnamefont {A.}~\bibnamefont {Matsushima}}, \bibinfo
  {author} {\bibfnamefont {J.~J.}\ \bibnamefont {Hudson}}, \bibinfo {author}
  {\bibfnamefont {E.}~\bibnamefont {Hinds}}, \bibinfo {author} {\bibfnamefont
  {M.}~\bibnamefont {Tarbutt}},\ and\ \bibinfo {author} {\bibfnamefont
  {B.}~\bibnamefont {Sauer}},\ }\href@noop {} {\bibfield  {journal} {\bibinfo
  {journal} {Phys. Rev. A}\ }\textbf {\bibinfo {volume} {89}},\ \bibinfo
  {pages} {053416} (\bibinfo {year} {2014})}\BibitemShut {NoStop}%
\bibitem [{\citenamefont {Shuman}\ \emph {et~al.}(2010)\citenamefont {Shuman},
  \citenamefont {Barry},\ and\ \citenamefont {DeMille}}]{shuman2010laser}%
  \BibitemOpen
  \bibfield  {author} {\bibinfo {author} {\bibfnamefont {E.~S.}\ \bibnamefont
  {Shuman}}, \bibinfo {author} {\bibfnamefont {J.~F.}\ \bibnamefont {Barry}},\
  and\ \bibinfo {author} {\bibfnamefont {D.}~\bibnamefont {DeMille}},\
  }\href@noop {} {\bibfield  {journal} {\bibinfo  {journal} {Nature}\ }\textbf
  {\bibinfo {volume} {467}},\ \bibinfo {pages} {820} (\bibinfo {year}
  {2010})}\BibitemShut {NoStop}%
\bibitem [{\citenamefont {Rockenh{\"a}user}\ \emph {et~al.}(2024)\citenamefont
  {Rockenh{\"a}user}, \citenamefont {Kogel}, \citenamefont {Garg},
  \citenamefont {Morales-Ram{\'\i}rez},\ and\ \citenamefont
  {Langen}}]{rockenhauser2024laser}%
  \BibitemOpen
  \bibfield  {author} {\bibinfo {author} {\bibfnamefont {M.}~\bibnamefont
  {Rockenh{\"a}user}}, \bibinfo {author} {\bibfnamefont {F.}~\bibnamefont
  {Kogel}}, \bibinfo {author} {\bibfnamefont {T.}~\bibnamefont {Garg}},
  \bibinfo {author} {\bibfnamefont {S.~A.}\ \bibnamefont
  {Morales-Ram{\'\i}rez}},\ and\ \bibinfo {author} {\bibfnamefont
  {T.}~\bibnamefont {Langen}},\ }\href@noop {} {\bibfield  {journal} {\bibinfo
  {journal} {Phys. Rev. Res.}\ }\textbf {\bibinfo {volume} {6}},\ \bibinfo
  {pages} {043161} (\bibinfo {year} {2024})}\BibitemShut {NoStop}%
\bibitem [{\citenamefont {Collopy}\ \emph {et~al.}(2018)\citenamefont
  {Collopy}, \citenamefont {Ding}, \citenamefont {Wu}, \citenamefont
  {Finneran}, \citenamefont {Anderegg}, \citenamefont {Augenbraun},
  \citenamefont {Doyle},\ and\ \citenamefont {Ye}}]{collopy20183d}%
  \BibitemOpen
  \bibfield  {author} {\bibinfo {author} {\bibfnamefont {A.~L.}\ \bibnamefont
  {Collopy}}, \bibinfo {author} {\bibfnamefont {S.}~\bibnamefont {Ding}},
  \bibinfo {author} {\bibfnamefont {Y.}~\bibnamefont {Wu}}, \bibinfo {author}
  {\bibfnamefont {I.~A.}\ \bibnamefont {Finneran}}, \bibinfo {author}
  {\bibfnamefont {L.}~\bibnamefont {Anderegg}}, \bibinfo {author}
  {\bibfnamefont {B.~L.}\ \bibnamefont {Augenbraun}}, \bibinfo {author}
  {\bibfnamefont {J.~M.}\ \bibnamefont {Doyle}},\ and\ \bibinfo {author}
  {\bibfnamefont {J.}~\bibnamefont {Ye}},\ }\href@noop {} {\bibfield  {journal}
  {\bibinfo  {journal} {Phys. Rev. Lett.}\ }\textbf {\bibinfo {volume} {121}},\
  \bibinfo {pages} {213201} (\bibinfo {year} {2018})}\BibitemShut {NoStop}%
\bibitem [{\citenamefont {Dunham}(1932)}]{dunham:32}%
  \BibitemOpen
  \bibfield  {author} {\bibinfo {author} {\bibfnamefont {J.}~\bibnamefont
  {Dunham}},\ }\href@noop {} {\bibfield  {journal} {\bibinfo  {journal} {Phys.
  Rev.}\ }\textbf {\bibinfo {volume} {41}},\ \bibinfo {pages} {721} (\bibinfo
  {year} {1932})}\BibitemShut {NoStop}%
\bibitem [{\citenamefont {Pekeris}(1934)}]{pekeris:34}%
  \BibitemOpen
  \bibfield  {author} {\bibinfo {author} {\bibfnamefont {C.~L.}\ \bibnamefont
  {Pekeris}},\ }\href {https://doi.org/10.1103/PhysRev.45.98} {\bibfield
  {journal} {\bibinfo  {journal} {Phys. Rev.}\ }\textbf {\bibinfo {volume}
  {45}},\ \bibinfo {pages} {98} (\bibinfo {year} {1934})}\BibitemShut {NoStop}%
\bibitem [{\citenamefont {Kratzer}(1920)}]{kratzer:20}%
  \BibitemOpen
  \bibfield  {author} {\bibinfo {author} {\bibfnamefont {A.}~\bibnamefont
  {Kratzer}},\ }\href@noop {} {\bibfield  {journal} {\bibinfo  {journal}
  {Zeits. f. Physik}\ }\textbf {\bibinfo {volume} {3}},\ \bibinfo {pages} {289}
  (\bibinfo {year} {1920})}\BibitemShut {NoStop}%
\bibitem [{\citenamefont {Burkhardt}\ and\ \citenamefont
  {Leventhal}(2007)}]{burkhardt:07}%
  \BibitemOpen
  \bibfield  {author} {\bibinfo {author} {\bibfnamefont {C.~E.}\ \bibnamefont
  {Burkhardt}}\ and\ \bibinfo {author} {\bibfnamefont {J.~J.}\ \bibnamefont
  {Leventhal}},\ }\href {https://doi.org/10.1119/1.2750377} {\bibfield
  {journal} {\bibinfo  {journal} {Am. J. Phys.}\ }\textbf {\bibinfo {volume}
  {75}},\ \bibinfo {pages} {686} (\bibinfo {year} {2007})}\BibitemShut
  {NoStop}%
\bibitem [{\citenamefont {Docenko}\ \emph {et~al.}(2004)\citenamefont
  {Docenko}, \citenamefont {Tamanis}, \citenamefont {Ferber}, \citenamefont
  {Pashov}, \citenamefont {Kn{\"o}ckel},\ and\ \citenamefont
  {Tiemann}}]{docenko:04}%
  \BibitemOpen
  \bibfield  {author} {\bibinfo {author} {\bibfnamefont {O.}~\bibnamefont
  {Docenko}}, \bibinfo {author} {\bibfnamefont {M.}~\bibnamefont {Tamanis}},
  \bibinfo {author} {\bibfnamefont {R.}~\bibnamefont {Ferber}}, \bibinfo
  {author} {\bibfnamefont {A.}~\bibnamefont {Pashov}}, \bibinfo {author}
  {\bibfnamefont {H.}~\bibnamefont {Kn{\"o}ckel}},\ and\ \bibinfo {author}
  {\bibfnamefont {E.}~\bibnamefont {Tiemann}},\ }\href@noop {} {\bibfield
  {journal} {\bibinfo  {journal} {Eur. Phys. J. D}\ }\textbf {\bibinfo {volume}
  {31}},\ \bibinfo {pages} {205} (\bibinfo {year} {2004})}\BibitemShut
  {NoStop}%
\bibitem [{\citenamefont {Docenko}\ \emph {et~al.}()\citenamefont {Docenko},
  \citenamefont {Tamanis}, \citenamefont {Ferber}, \citenamefont
  {Kn{\"o}ckel},\ and\ \citenamefont {Tiemann}}]{docenko:11}%
  \BibitemOpen
  \bibfield  {author} {\bibinfo {author} {\bibfnamefont {O.}~\bibnamefont
  {Docenko}}, \bibinfo {author} {\bibfnamefont {M.}~\bibnamefont {Tamanis}},
  \bibinfo {author} {\bibfnamefont {R.}~\bibnamefont {Ferber}}, \bibinfo
  {author} {\bibfnamefont {H.}~\bibnamefont {Kn{\"o}ckel}},\ and\ \bibinfo
  {author} {\bibfnamefont {E.}~\bibnamefont {Tiemann}},\ }\href@noop {}
  {\bibinfo  {journal} {Phys. Rev. A}\ }\BibitemShut {NoStop}%
\bibitem [{\citenamefont {Pashov}\ \emph {et~al.}(2005)\citenamefont {Pashov},
  \citenamefont {Docenko}, \citenamefont {Tamanis}, \citenamefont {Ferber},
  \citenamefont {Kn\"ockel},\ and\ \citenamefont {Tiemann}}]{pashov:05}%
  \BibitemOpen
\bibfield  {journal} {  }\bibfield  {author} {\bibinfo {author} {\bibfnamefont
  {A.}~\bibnamefont {Pashov}}, \bibinfo {author} {\bibfnamefont
  {O.}~\bibnamefont {Docenko}}, \bibinfo {author} {\bibfnamefont
  {M.}~\bibnamefont {Tamanis}}, \bibinfo {author} {\bibfnamefont
  {R.}~\bibnamefont {Ferber}}, \bibinfo {author} {\bibfnamefont
  {H.}~\bibnamefont {Kn\"ockel}},\ and\ \bibinfo {author} {\bibfnamefont
  {E.}~\bibnamefont {Tiemann}},\ }\href
  {https://doi.org/10.1103/PhysRevA.72.062505} {\bibfield  {journal} {\bibinfo
  {journal} {Phys. Rev. A}\ }\textbf {\bibinfo {volume} {72}},\ \bibinfo
  {pages} {062505} (\bibinfo {year} {2005})}\BibitemShut {NoStop}%
\bibitem [{\citenamefont {Pashov}\ \emph {et~al.}(2007)\citenamefont {Pashov},
  \citenamefont {Docenko}, \citenamefont {Tamanis}, \citenamefont {Ferber},
  \citenamefont {Kn\"ockel},\ and\ \citenamefont {Tiemann}}]{pashov:07}%
  \BibitemOpen
  \bibfield  {author} {\bibinfo {author} {\bibfnamefont {A.}~\bibnamefont
  {Pashov}}, \bibinfo {author} {\bibfnamefont {O.}~\bibnamefont {Docenko}},
  \bibinfo {author} {\bibfnamefont {M.}~\bibnamefont {Tamanis}}, \bibinfo
  {author} {\bibfnamefont {R.}~\bibnamefont {Ferber}}, \bibinfo {author}
  {\bibfnamefont {H.}~\bibnamefont {Kn\"ockel}},\ and\ \bibinfo {author}
  {\bibfnamefont {E.}~\bibnamefont {Tiemann}},\ }\href
  {https://doi.org/10.1103/PhysRevA.76.022511} {\bibfield  {journal} {\bibinfo
  {journal} {Phys. Rev. A}\ }\textbf {\bibinfo {volume} {76}},\ \bibinfo
  {pages} {022511} (\bibinfo {year} {2007})}\BibitemShut {NoStop}%
\bibitem [{\citenamefont {Steinke}\ \emph {et~al.}(2012)\citenamefont
  {Steinke}, \citenamefont {Kn\"ockel},\ and\ \citenamefont
  {Tiemann}}]{steinke:12}%
  \BibitemOpen
  \bibfield  {author} {\bibinfo {author} {\bibfnamefont {M.}~\bibnamefont
  {Steinke}}, \bibinfo {author} {\bibfnamefont {H.}~\bibnamefont {Kn\"ockel}},\
  and\ \bibinfo {author} {\bibfnamefont {E.}~\bibnamefont {Tiemann}},\ }\href
  {https://doi.org/10.1103/PhysRevA.85.042720} {\bibfield  {journal} {\bibinfo
  {journal} {Phys. Rev. A}\ }\textbf {\bibinfo {volume} {85}},\ \bibinfo
  {pages} {042720} (\bibinfo {year} {2012})}\BibitemShut {NoStop}%
\bibitem [{\citenamefont {Ferber}\ \emph {et~al.}(2009)\citenamefont {Ferber},
  \citenamefont {Klincare}, \citenamefont {Nikolayeva}, \citenamefont
  {Tamanis}, \citenamefont {Kn{\"o}ckel}, \citenamefont {Tiemann},\ and\
  \citenamefont {Pashov}}]{ferber:09}%
  \BibitemOpen
  \bibfield  {author} {\bibinfo {author} {\bibfnamefont {R.}~\bibnamefont
  {Ferber}}, \bibinfo {author} {\bibfnamefont {I.}~\bibnamefont {Klincare}},
  \bibinfo {author} {\bibfnamefont {O.}~\bibnamefont {Nikolayeva}}, \bibinfo
  {author} {\bibfnamefont {M.}~\bibnamefont {Tamanis}}, \bibinfo {author}
  {\bibfnamefont {H.}~\bibnamefont {Kn{\"o}ckel}}, \bibinfo {author}
  {\bibfnamefont {E.}~\bibnamefont {Tiemann}},\ and\ \bibinfo {author}
  {\bibfnamefont {A.}~\bibnamefont {Pashov}},\ }\href@noop {} {\bibfield
  {journal} {\bibinfo  {journal} {Phys. Rev. A}\ }\textbf {\bibinfo {volume}
  {80}},\ \bibinfo {pages} {062501} (\bibinfo {year} {2009})}\BibitemShut
  {NoStop}%
\bibitem [{\citenamefont {Ivanova}\ \emph {et~al.}(2011)\citenamefont
  {Ivanova}, \citenamefont {Stein}, \citenamefont {Pashov}, \citenamefont
  {Knöckel},\ and\ \citenamefont {Tiemann}}]{ivanova:11}%
  \BibitemOpen
  \bibfield  {author} {\bibinfo {author} {\bibfnamefont {M.}~\bibnamefont
  {Ivanova}}, \bibinfo {author} {\bibfnamefont {A.}~\bibnamefont {Stein}},
  \bibinfo {author} {\bibfnamefont {A.}~\bibnamefont {Pashov}}, \bibinfo
  {author} {\bibfnamefont {H.}~\bibnamefont {Knöckel}},\ and\ \bibinfo
  {author} {\bibfnamefont {E.}~\bibnamefont {Tiemann}},\ }\href
  {https://doi.org/10.1063/1.3524312} {\bibfield  {journal} {\bibinfo
  {journal} {J. Chem. Phys.}\ }\textbf {\bibinfo {volume} {134}},\ \bibinfo
  {pages} {024321} (\bibinfo {year} {2011})}\BibitemShut {NoStop}%
\bibitem [{\citenamefont {Tiemann}\ \emph {et~al.}(2009)\citenamefont
  {Tiemann}, \citenamefont {Kn{\"o}ckel}, \citenamefont {Kowalczyk},
  \citenamefont {Jastrzebski}, \citenamefont {Pashov}, \citenamefont {Salami},\
  and\ \citenamefont {Ross}}]{tiemann:09}%
  \BibitemOpen
  \bibfield  {author} {\bibinfo {author} {\bibfnamefont {E.}~\bibnamefont
  {Tiemann}}, \bibinfo {author} {\bibfnamefont {H.}~\bibnamefont
  {Kn{\"o}ckel}}, \bibinfo {author} {\bibfnamefont {P.}~\bibnamefont
  {Kowalczyk}}, \bibinfo {author} {\bibfnamefont {W.}~\bibnamefont
  {Jastrzebski}}, \bibinfo {author} {\bibfnamefont {A.}~\bibnamefont {Pashov}},
  \bibinfo {author} {\bibfnamefont {H.}~\bibnamefont {Salami}},\ and\ \bibinfo
  {author} {\bibfnamefont {A.}~\bibnamefont {Ross}},\ }\href@noop {} {\bibfield
   {journal} {\bibinfo  {journal} {Phys. Rev. A}\ }\textbf {\bibinfo {volume}
  {79}},\ \bibinfo {pages} {042716} (\bibinfo {year} {2009})}\BibitemShut
  {NoStop}%
\bibitem [{\citenamefont {\ifmmode~\acute{S}\else \'{S}\fi{}mia\l{}kowski}\
  and\ \citenamefont {Tomza}(2021)}]{Smialkowski}%
  \BibitemOpen
  \bibfield  {author} {\bibinfo {author} {\bibfnamefont {M.}~\bibnamefont
  {\ifmmode~\acute{S}\else \'{S}\fi{}mia\l{}kowski}}\ and\ \bibinfo {author}
  {\bibfnamefont {M.}~\bibnamefont {Tomza}},\ }\href@noop {} {\bibfield
  {journal} {\bibinfo  {journal} {Phys. Rev. A}\ }\textbf {\bibinfo {volume}
  {103}},\ \bibinfo {pages} {022802} (\bibinfo {year} {2021})}\BibitemShut
  {NoStop}%
\bibitem [{\citenamefont {Aymar}\ and\ \citenamefont
  {Dulieu}(2005)}]{aymar2005calculation}%
  \BibitemOpen
  \bibfield  {author} {\bibinfo {author} {\bibfnamefont {M.}~\bibnamefont
  {Aymar}}\ and\ \bibinfo {author} {\bibfnamefont {O.}~\bibnamefont {Dulieu}},\
  }\href@noop {} {\bibfield  {journal} {\bibinfo  {journal} {J. Chem. Phys.}\
  }\textbf {\bibinfo {volume} {122}} (\bibinfo {year} {2005})}\BibitemShut
  {NoStop}%
\bibitem [{\citenamefont {Ladjimi}\ and\ \citenamefont
  {Tomza}(2024)}]{Ladjimi}%
  \BibitemOpen
  \bibfield  {author} {\bibinfo {author} {\bibfnamefont {H.}~\bibnamefont
  {Ladjimi}}\ and\ \bibinfo {author} {\bibfnamefont {M.}~\bibnamefont
  {Tomza}},\ }\href {https://doi.org/10.1103/PhysRevA.109.052814} {\bibfield
  {journal} {\bibinfo  {journal} {Phys. Rev. A}\ }\textbf {\bibinfo {volume}
  {109}},\ \bibinfo {pages} {052814} (\bibinfo {year} {2024})}\BibitemShut
  {NoStop}%
\bibitem [{\citenamefont {Koput}(2015)}]{koput2015ab}%
  \BibitemOpen
  \bibfield  {author} {\bibinfo {author} {\bibfnamefont {J.}~\bibnamefont
  {Koput}},\ }\href@noop {} {\bibfield  {journal} {\bibinfo  {journal} {J.
  Comput. Chem.}\ }\textbf {\bibinfo {volume} {36}},\ \bibinfo {pages} {1286}
  (\bibinfo {year} {2015})}\BibitemShut {NoStop}%
\bibitem [{\citenamefont {Radford}\ and\ \citenamefont
  {Litvak}(1975)}]{radford1975imine}%
  \BibitemOpen
  \bibfield  {author} {\bibinfo {author} {\bibfnamefont {H.}~\bibnamefont
  {Radford}}\ and\ \bibinfo {author} {\bibfnamefont {M.}~\bibnamefont
  {Litvak}},\ }\href@noop {} {\bibfield  {journal} {\bibinfo  {journal} {Chem.
  Phys. Lett.}\ }\textbf {\bibinfo {volume} {34}},\ \bibinfo {pages} {561}
  (\bibinfo {year} {1975})}\BibitemShut {NoStop}%
\bibitem [{\citenamefont {Wayne}\ and\ \citenamefont
  {Radford}(1976)}]{wayne1976laser}%
  \BibitemOpen
  \bibfield  {author} {\bibinfo {author} {\bibfnamefont {F.}~\bibnamefont
  {Wayne}}\ and\ \bibinfo {author} {\bibfnamefont {H.}~\bibnamefont
  {Radford}},\ }\href@noop {} {\bibfield  {journal} {\bibinfo  {journal} {Mol.
  Phys.}\ }\textbf {\bibinfo {volume} {32}},\ \bibinfo {pages} {1407} (\bibinfo
  {year} {1976})}\BibitemShut {NoStop}%
\bibitem [{\citenamefont {T{\"o}rring}\ \emph {et~al.}(1984)\citenamefont
  {T{\"o}rring}, \citenamefont {Ernst},\ and\ \citenamefont
  {Kindt}}]{torring1984dipole}%
  \BibitemOpen
  \bibfield  {author} {\bibinfo {author} {\bibfnamefont {T.}~\bibnamefont
  {T{\"o}rring}}, \bibinfo {author} {\bibfnamefont {W.}~\bibnamefont {Ernst}},\
  and\ \bibinfo {author} {\bibfnamefont {S.}~\bibnamefont {Kindt}},\
  }\href@noop {} {\bibfield  {journal} {\bibinfo  {journal} {J. Chem. Phys.}\
  }\textbf {\bibinfo {volume} {81}},\ \bibinfo {pages} {4614} (\bibinfo {year}
  {1984})}\BibitemShut {NoStop}%
\bibitem [{pas()}]{pasteka}%
  \BibitemOpen
  \href@noop {} {}\bibinfo {note} {Private communication from Luk{\'a}{\v{s}}
  Pa{\v{s}}teka and Anastasia Borshevsky, based on MRCI calculations similar to
  those reported in J. Chem. Phys., {\bf 151} 034302 (2019), but using a
  smaller basis set.}\BibitemShut {Stop}%
\bibitem [{\citenamefont {Huber}\ and\ \citenamefont
  {Herzberg}(1979)}]{Huber_1979}%
  \BibitemOpen
  \bibfield  {author} {\bibinfo {author} {\bibfnamefont {K.~P.}\ \bibnamefont
  {Huber}}\ and\ \bibinfo {author} {\bibfnamefont {G.}~\bibnamefont
  {Herzberg}},\ }\href {https://doi.org/10.1007/978-1-4757-0961-2} {\emph
  {\bibinfo {title} {Molecular Spectra and Molecular Structure}}}\ (\bibinfo
  {publisher} {Springer US},\ \bibinfo {year} {1979})\BibitemShut {NoStop}%
\bibitem [{\citenamefont {Childs}\ \emph {et~al.}(1986)\citenamefont {Childs},
  \citenamefont {Goodman},\ and\ \citenamefont {Goodman}}]{CHILDS1986215}%
  \BibitemOpen
  \bibfield  {author} {\bibinfo {author} {\bibfnamefont {W.}~\bibnamefont
  {Childs}}, \bibinfo {author} {\bibfnamefont {G.}~\bibnamefont {Goodman}},\
  and\ \bibinfo {author} {\bibfnamefont {L.}~\bibnamefont {Goodman}},\ }\href
  {https://doi.org/10.1016/0022-2852(86)90288-2} {\bibfield  {journal}
  {\bibinfo  {journal} {J. Mol. Spectr.}\ }\textbf {\bibinfo {volume} {115}},\
  \bibinfo {pages} {215} (\bibinfo {year} {1986})}\BibitemShut {NoStop}%
\bibitem [{\citenamefont {Hao}\ \emph {et~al.}(2019)\citenamefont {Hao},
  \citenamefont {Pašteka}, \citenamefont {Visscher}, \citenamefont {Aggarwal},
  \citenamefont {Bethlem}, \citenamefont {Boeschoten}, \citenamefont
  {Borschevsky}, \citenamefont {Denis}, \citenamefont {Esajas}, \citenamefont
  {Hoekstra}, \citenamefont {Jungmann}, \citenamefont {Marshall}, \citenamefont
  {Meijknecht}, \citenamefont {Mooij}, \citenamefont {Timmermans},
  \citenamefont {Touwen}, \citenamefont {Ubachs}, \citenamefont {Willmann},
  \citenamefont {Yin}, \citenamefont {Zapara},\ and\ \citenamefont {eEDM
  Collaboration)}}]{hao}%
  \BibitemOpen
  \bibfield  {author} {\bibinfo {author} {\bibfnamefont {Y.}~\bibnamefont
  {Hao}}, \bibinfo {author} {\bibfnamefont {L.~F.}\ \bibnamefont {Pašteka}},
  \bibinfo {author} {\bibfnamefont {L.}~\bibnamefont {Visscher}}, \bibinfo
  {author} {\bibfnamefont {P.}~\bibnamefont {Aggarwal}}, \bibinfo {author}
  {\bibfnamefont {H.~L.}\ \bibnamefont {Bethlem}}, \bibinfo {author}
  {\bibfnamefont {A.}~\bibnamefont {Boeschoten}}, \bibinfo {author}
  {\bibfnamefont {A.}~\bibnamefont {Borschevsky}}, \bibinfo {author}
  {\bibfnamefont {M.}~\bibnamefont {Denis}}, \bibinfo {author} {\bibfnamefont
  {K.}~\bibnamefont {Esajas}}, \bibinfo {author} {\bibfnamefont
  {S.}~\bibnamefont {Hoekstra}}, \bibinfo {author} {\bibfnamefont
  {K.}~\bibnamefont {Jungmann}}, \bibinfo {author} {\bibfnamefont {V.~R.}\
  \bibnamefont {Marshall}}, \bibinfo {author} {\bibfnamefont {T.~B.}\
  \bibnamefont {Meijknecht}}, \bibinfo {author} {\bibfnamefont {M.~C.}\
  \bibnamefont {Mooij}}, \bibinfo {author} {\bibfnamefont {R.~G.~E.}\
  \bibnamefont {Timmermans}}, \bibinfo {author} {\bibfnamefont
  {A.}~\bibnamefont {Touwen}}, \bibinfo {author} {\bibfnamefont
  {W.}~\bibnamefont {Ubachs}}, \bibinfo {author} {\bibfnamefont
  {L.}~\bibnamefont {Willmann}}, \bibinfo {author} {\bibfnamefont
  {Y.}~\bibnamefont {Yin}}, \bibinfo {author} {\bibfnamefont {A.}~\bibnamefont
  {Zapara}},\ and\ \bibinfo {author} {\bibfnamefont {N.}~\bibnamefont {eEDM
  Collaboration)}},\ }\href {https://doi.org/10.1063/1.5098540} {\bibfield
  {journal} {\bibinfo  {journal} {J. Chem. Phys.}\ }\textbf {\bibinfo {volume}
  {151}},\ \bibinfo {pages} {034302} (\bibinfo {year} {2019})}\BibitemShut
  {NoStop}%
\bibitem [{\citenamefont {{Anderson}}\ \emph {et~al.}(1994)\citenamefont
  {{Anderson}}, \citenamefont {{Allen}},\ and\ \citenamefont
  {{Ziurys}}}]{anderson:94}%
  \BibitemOpen
  \bibfield  {author} {\bibinfo {author} {\bibfnamefont {M.~A.}\ \bibnamefont
  {{Anderson}}}, \bibinfo {author} {\bibfnamefont {M.~D.}\ \bibnamefont
  {{Allen}}},\ and\ \bibinfo {author} {\bibfnamefont {L.~M.}\ \bibnamefont
  {{Ziurys}}},\ }\href {https://doi.org/10.1086/173907} {\bibfield  {journal}
  {\bibinfo  {journal} {Astrophys. J.}\ }\textbf {\bibinfo {volume} {424}},\
  \bibinfo {pages} {503} (\bibinfo {year} {1994})}\BibitemShut {NoStop}%
\bibitem [{\citenamefont {Ernst}\ \emph {et~al.}(1985)\citenamefont {Ernst},
  \citenamefont {K{\"a}ndler}, \citenamefont {Kindt},\ and\ \citenamefont
  {T{\"o}rring}}]{ernst1985electric}%
  \BibitemOpen
  \bibfield  {author} {\bibinfo {author} {\bibfnamefont {W.}~\bibnamefont
  {Ernst}}, \bibinfo {author} {\bibfnamefont {J.}~\bibnamefont {K{\"a}ndler}},
  \bibinfo {author} {\bibfnamefont {S.}~\bibnamefont {Kindt}},\ and\ \bibinfo
  {author} {\bibfnamefont {T.}~\bibnamefont {T{\"o}rring}},\ }\href@noop {}
  {\bibfield  {journal} {\bibinfo  {journal} {Chem. Phys. Lett.}\ }\textbf
  {\bibinfo {volume} {113}},\ \bibinfo {pages} {351} (\bibinfo {year}
  {1985})}\BibitemShut {NoStop}%
\bibitem [{\citenamefont {Suenram}\ \emph {et~al.}(1990)\citenamefont
  {Suenram}, \citenamefont {Lovas}, \citenamefont {Fraser},\ and\ \citenamefont
  {Matsumura}}]{suenram}%
  \BibitemOpen
  \bibfield  {author} {\bibinfo {author} {\bibfnamefont {R.~D.}\ \bibnamefont
  {Suenram}}, \bibinfo {author} {\bibfnamefont {F.~J.}\ \bibnamefont {Lovas}},
  \bibinfo {author} {\bibfnamefont {G.~T.}\ \bibnamefont {Fraser}},\ and\
  \bibinfo {author} {\bibfnamefont {K.}~\bibnamefont {Matsumura}},\ }\href
  {https://doi.org/10.1063/1.457690} {\bibfield  {journal} {\bibinfo  {journal}
  {J. Chem. Phys.}\ }\textbf {\bibinfo {volume} {92}},\ \bibinfo {pages} {4724}
  (\bibinfo {year} {1990})}\BibitemShut {NoStop}%
\bibitem [{\citenamefont {Staanum}\ \emph {et~al.}(2007)\citenamefont
  {Staanum}, \citenamefont {Pashov}, \citenamefont {Kn{\"o}ckel},\ and\
  \citenamefont {Tiemann}}]{staanum:07}%
  \BibitemOpen
  \bibfield  {author} {\bibinfo {author} {\bibfnamefont {P.}~\bibnamefont
  {Staanum}}, \bibinfo {author} {\bibfnamefont {A.}~\bibnamefont {Pashov}},
  \bibinfo {author} {\bibfnamefont {H.}~\bibnamefont {Kn{\"o}ckel}},\ and\
  \bibinfo {author} {\bibfnamefont {E.}~\bibnamefont {Tiemann}},\ }\href@noop
  {} {\bibfield  {journal} {\bibinfo  {journal} {Phys. Rev. A}\ }\textbf
  {\bibinfo {volume} {75}},\ \bibinfo {pages} {042513} (\bibinfo {year}
  {2007})}\BibitemShut {NoStop}%
\bibitem [{\citenamefont {Hartmann}\ \emph {et~al.}(2019)\citenamefont
  {Hartmann}, \citenamefont {Schulze}, \citenamefont {Voges}, \citenamefont
  {Gersema}, \citenamefont {Gempel}, \citenamefont {Tiemann}, \citenamefont
  {Zenesini},\ and\ \citenamefont {Ospelkaus}}]{hartmann:19}%
  \BibitemOpen
  \bibfield  {author} {\bibinfo {author} {\bibfnamefont {T.}~\bibnamefont
  {Hartmann}}, \bibinfo {author} {\bibfnamefont {T.~A.}\ \bibnamefont
  {Schulze}}, \bibinfo {author} {\bibfnamefont {K.~K.}\ \bibnamefont {Voges}},
  \bibinfo {author} {\bibfnamefont {P.}~\bibnamefont {Gersema}}, \bibinfo
  {author} {\bibfnamefont {M.~W.}\ \bibnamefont {Gempel}}, \bibinfo {author}
  {\bibfnamefont {E.}~\bibnamefont {Tiemann}}, \bibinfo {author} {\bibfnamefont
  {A.}~\bibnamefont {Zenesini}},\ and\ \bibinfo {author} {\bibfnamefont
  {S.}~\bibnamefont {Ospelkaus}},\ }\href@noop {} {\bibfield  {journal}
  {\bibinfo  {journal} {Phys. Rev. A}\ }\textbf {\bibinfo {volume} {99}},\
  \bibinfo {pages} {032711} (\bibinfo {year} {2019})}\BibitemShut {NoStop}%
\bibitem [{\citenamefont {Idziaszek}\ and\ \citenamefont
  {Julienne}(2010)}]{idziaszek2010universal}%
  \BibitemOpen
  \bibfield  {author} {\bibinfo {author} {\bibfnamefont {Z.}~\bibnamefont
  {Idziaszek}}\ and\ \bibinfo {author} {\bibfnamefont {P.~S.}\ \bibnamefont
  {Julienne}},\ }\href@noop {} {\bibfield  {journal} {\bibinfo  {journal}
  {Phys. Rev. Lett.}\ }\textbf {\bibinfo {volume} {104}},\ \bibinfo {pages}
  {113202} (\bibinfo {year} {2010})}\BibitemShut {NoStop}%
\bibitem [{\citenamefont {Karman}\ and\ \citenamefont
  {Hutson}(2018)}]{karman2018microwave}%
  \BibitemOpen
  \bibfield  {author} {\bibinfo {author} {\bibfnamefont {T.}~\bibnamefont
  {Karman}}\ and\ \bibinfo {author} {\bibfnamefont {J.~M.}\ \bibnamefont
  {Hutson}},\ }\href@noop {} {\bibfield  {journal} {\bibinfo  {journal} {Phys.
  Rev. Lett.}\ }\textbf {\bibinfo {volume} {121}},\ \bibinfo {pages} {163401}
  (\bibinfo {year} {2018})}\BibitemShut {NoStop}%
\bibitem [{\citenamefont {Anderegg}\ \emph {et~al.}(2021)\citenamefont
  {Anderegg}, \citenamefont {Burchesky}, \citenamefont {Bao}, \citenamefont
  {Yu}, \citenamefont {Karman}, \citenamefont {Chae}, \citenamefont {Ni},
  \citenamefont {Ketterle},\ and\ \citenamefont {Doyle}}]{anderegg:21}%
  \BibitemOpen
  \bibfield  {author} {\bibinfo {author} {\bibfnamefont {L.}~\bibnamefont
  {Anderegg}}, \bibinfo {author} {\bibfnamefont {S.}~\bibnamefont {Burchesky}},
  \bibinfo {author} {\bibfnamefont {Y.}~\bibnamefont {Bao}}, \bibinfo {author}
  {\bibfnamefont {S.~S.}\ \bibnamefont {Yu}}, \bibinfo {author} {\bibfnamefont
  {T.}~\bibnamefont {Karman}}, \bibinfo {author} {\bibfnamefont
  {E.}~\bibnamefont {Chae}}, \bibinfo {author} {\bibfnamefont {K.-K.}\
  \bibnamefont {Ni}}, \bibinfo {author} {\bibfnamefont {W.}~\bibnamefont
  {Ketterle}},\ and\ \bibinfo {author} {\bibfnamefont {J.~M.}\ \bibnamefont
  {Doyle}},\ }\href {https://doi.org/10.1126/science.abg9502} {\bibfield
  {journal} {\bibinfo  {journal} {Science}\ }\textbf {\bibinfo {volume}
  {373}},\ \bibinfo {pages} {779} (\bibinfo {year} {2021})}\BibitemShut
  {NoStop}%
\bibitem [{\citenamefont {Bigagli}\ \emph {et~al.}(2023)\citenamefont
  {Bigagli}, \citenamefont {Warner}, \citenamefont {Yuan}, \citenamefont
  {Zhang}, \citenamefont {Stevenson}, \citenamefont {Karman},\ and\
  \citenamefont {Will}}]{bigagli:23}%
  \BibitemOpen
  \bibfield  {author} {\bibinfo {author} {\bibfnamefont {N.}~\bibnamefont
  {Bigagli}}, \bibinfo {author} {\bibfnamefont {C.}~\bibnamefont {Warner}},
  \bibinfo {author} {\bibfnamefont {W.}~\bibnamefont {Yuan}}, \bibinfo {author}
  {\bibfnamefont {S.}~\bibnamefont {Zhang}}, \bibinfo {author} {\bibfnamefont
  {I.}~\bibnamefont {Stevenson}}, \bibinfo {author} {\bibfnamefont
  {T.}~\bibnamefont {Karman}},\ and\ \bibinfo {author} {\bibfnamefont
  {S.}~\bibnamefont {Will}},\ }\href
  {https://doi.org/10.1038/s41567-023-02200-6} {\bibfield  {journal} {\bibinfo
  {journal} {Nature Phys.}\ }\textbf {\bibinfo {volume} {19}},\ \bibinfo
  {pages} {1579} (\bibinfo {year} {2023})}\BibitemShut {NoStop}%
\bibitem [{\citenamefont {Janssen}\ \emph {et~al.}(2013)\citenamefont
  {Janssen}, \citenamefont {van~der Avoird},\ and\ \citenamefont
  {Groenenboom}}]{janssen2013quantum}%
  \BibitemOpen
  \bibfield  {author} {\bibinfo {author} {\bibfnamefont {L.~M.}\ \bibnamefont
  {Janssen}}, \bibinfo {author} {\bibfnamefont {A.}~\bibnamefont {van~der
  Avoird}},\ and\ \bibinfo {author} {\bibfnamefont {G.~C.}\ \bibnamefont
  {Groenenboom}},\ }\href@noop {} {\bibfield  {journal} {\bibinfo  {journal}
  {Phys. Rev. Lett.}\ }\textbf {\bibinfo {volume} {110}},\ \bibinfo {pages}
  {063201} (\bibinfo {year} {2013})}\BibitemShut {NoStop}%
\bibitem [{\citenamefont {Karman}(2023)}]{karman:23}%
  \BibitemOpen
  \bibfield  {author} {\bibinfo {author} {\bibfnamefont {T.}~\bibnamefont
  {Karman}},\ }\href {https://doi.org/10.1021/acs.jpca.3c00797} {\bibfield
  {journal} {\bibinfo  {journal} {J. Phys. Chem. A}\ }\textbf {\bibinfo
  {volume} {127}},\ \bibinfo {pages} {2194} (\bibinfo {year} {2023})},\ \Eprint
  {https://arxiv.org/abs/arXiv:2212.03065} {arXiv:2212.03065} \BibitemShut
  {NoStop}%
\bibitem [{\citenamefont {P{\'e}rez}\ \emph {et~al.}()\citenamefont
  {P{\'e}rez}, \citenamefont {Desroches}, \citenamefont {Hazzard},\ and\
  \citenamefont {Karman}}]{kevin}%
  \BibitemOpen
  \bibfield  {author} {\bibinfo {author} {\bibfnamefont {K.}~\bibnamefont
  {P{\'e}rez}}, \bibinfo {author} {\bibfnamefont {J.~W.}\ \bibnamefont
  {Desroches}}, \bibinfo {author} {\bibfnamefont {K.~R.~A.}\ \bibnamefont
  {Hazzard}},\ and\ \bibinfo {author} {\bibfnamefont {T.}~\bibnamefont
  {Karman}},\ }\href@noop {} {\bibinfo {title} {Hubbard models with ultracold
  shielded molecules in optical lattices}},\ \bibinfo {note} {in
  preparation}\BibitemShut {NoStop}%
\bibitem [{\citenamefont {Duda}\ \emph {et~al.}(2023)\citenamefont {Duda},
  \citenamefont {Chen}, \citenamefont {Schindewolf}, \citenamefont {Bause},
  \citenamefont {von Milczewski}, \citenamefont {Schmidt}, \citenamefont
  {Bloch},\ and\ \citenamefont {Luo}}]{duda:23}%
  \BibitemOpen
  \bibfield  {author} {\bibinfo {author} {\bibfnamefont {M.}~\bibnamefont
  {Duda}}, \bibinfo {author} {\bibfnamefont {X.-Y.}\ \bibnamefont {Chen}},
  \bibinfo {author} {\bibfnamefont {A.}~\bibnamefont {Schindewolf}}, \bibinfo
  {author} {\bibfnamefont {R.}~\bibnamefont {Bause}}, \bibinfo {author}
  {\bibfnamefont {J.}~\bibnamefont {von Milczewski}}, \bibinfo {author}
  {\bibfnamefont {R.}~\bibnamefont {Schmidt}}, \bibinfo {author} {\bibfnamefont
  {I.}~\bibnamefont {Bloch}},\ and\ \bibinfo {author} {\bibfnamefont {X.-Y.}\
  \bibnamefont {Luo}},\ }\href@noop {} {\bibfield  {journal} {\bibinfo
  {journal} {Nature Physics}\ }\textbf {\bibinfo {volume} {19}},\ \bibinfo
  {pages} {720} (\bibinfo {year} {2023})}\BibitemShut {NoStop}%
\bibitem [{\citenamefont {Li}\ \emph {et~al.}(2021)\citenamefont {Li},
  \citenamefont {Tobias}, \citenamefont {Matsuda}, \citenamefont {Miller},
  \citenamefont {Valtolina}, \citenamefont {De~Marco}, \citenamefont {Wang},
  \citenamefont {Lassabli{\`e}re}, \citenamefont {Qu{\'e}m{\'e}ner},
  \citenamefont {Bohn} \emph {et~al.}}]{li:21}%
  \BibitemOpen
  \bibfield  {author} {\bibinfo {author} {\bibfnamefont {J.-R.}\ \bibnamefont
  {Li}}, \bibinfo {author} {\bibfnamefont {W.~G.}\ \bibnamefont {Tobias}},
  \bibinfo {author} {\bibfnamefont {K.}~\bibnamefont {Matsuda}}, \bibinfo
  {author} {\bibfnamefont {C.}~\bibnamefont {Miller}}, \bibinfo {author}
  {\bibfnamefont {G.}~\bibnamefont {Valtolina}}, \bibinfo {author}
  {\bibfnamefont {L.}~\bibnamefont {De~Marco}}, \bibinfo {author}
  {\bibfnamefont {R.~R.}\ \bibnamefont {Wang}}, \bibinfo {author}
  {\bibfnamefont {L.}~\bibnamefont {Lassabli{\`e}re}}, \bibinfo {author}
  {\bibfnamefont {G.}~\bibnamefont {Qu{\'e}m{\'e}ner}}, \bibinfo {author}
  {\bibfnamefont {J.~L.}\ \bibnamefont {Bohn}}, \emph {et~al.},\ }\href
  {https://doi.org/10.1038/s41567-021-01329-6} {\bibfield  {journal} {\bibinfo
  {journal} {Nature Phys.}\ }\textbf {\bibinfo {volume} {17}},\ \bibinfo
  {pages} {1144} (\bibinfo {year} {2021})}\BibitemShut {NoStop}%
\bibitem [{\citenamefont {J{\'o}{\'z}wiak}\ \emph {et~al.}(2026)\citenamefont
  {J{\'o}{\'z}wiak}, \citenamefont {Yang}, \citenamefont {Dizer}, \citenamefont
  {Christianen},\ and\ \citenamefont {Karman}}]{spin}%
  \BibitemOpen
  \bibfield  {author} {\bibinfo {author} {\bibfnamefont {H.~J.}\ \bibnamefont
  {J{\'o}{\'z}wiak}}, \bibinfo {author} {\bibfnamefont {H.}~\bibnamefont
  {Yang}}, \bibinfo {author} {\bibfnamefont {E.}~\bibnamefont {Dizer}},
  \bibinfo {author} {\bibfnamefont {A.}~\bibnamefont {Christianen}},\ and\
  \bibinfo {author} {\bibfnamefont {T.}~\bibnamefont {Karman}},\ }\href
  {https://arxiv.org/abs/2607.25777} {\bibinfo {title} {Tunable state-dependent
  interactions in collisionally stable mixtures of polar molecules}} (\bibinfo
  {year} {2026}),\ \Eprint {https://arxiv.org/abs/2607.25777} {arXiv:2607.25777
  [cond-mat.quant-gas]} \BibitemShut {NoStop}%
\bibitem [{\citenamefont {Saffman}\ and\ \citenamefont
  {M{\o}lmer}(2009)}]{saffman2009efficient}%
  \BibitemOpen
  \bibfield  {author} {\bibinfo {author} {\bibfnamefont {M.}~\bibnamefont
  {Saffman}}\ and\ \bibinfo {author} {\bibfnamefont {K.}~\bibnamefont
  {M{\o}lmer}},\ }\href@noop {} {\bibfield  {journal} {\bibinfo  {journal}
  {Phys. Rev. Lett.}\ }\textbf {\bibinfo {volume} {102}},\ \bibinfo {pages}
  {240502} (\bibinfo {year} {2009})}\BibitemShut {NoStop}%
\bibitem [{\citenamefont {Yang}\ \emph {et~al.}()\citenamefont {Yang},
  \citenamefont {Dizer}, \citenamefont {J\'{o}\'{z}wiak}, \citenamefont
  {Schmidt}, \citenamefont {Christianen},\ and\ \citenamefont
  {Karman}}]{polaron}%
  \BibitemOpen
  \bibfield  {author} {\bibinfo {author} {\bibfnamefont {H.}~\bibnamefont
  {Yang}}, \bibinfo {author} {\bibfnamefont {E.}~\bibnamefont {Dizer}},
  \bibinfo {author} {\bibfnamefont {H.~J.}\ \bibnamefont {J\'{o}\'{z}wiak}},
  \bibinfo {author} {\bibfnamefont {R.}~\bibnamefont {Schmidt}}, \bibinfo
  {author} {\bibfnamefont {A.}~\bibnamefont {Christianen}},\ and\ \bibinfo
  {author} {\bibfnamefont {T.}~\bibnamefont {Karman}},\ }\href@noop {}
  {\bibinfo {title} {{Polarons in Microwave-Shielded Vibrational Mixtures of
  Ultracold Molecules}}},\ \bibinfo {note} {in preparation}\BibitemShut
  {NoStop}%
\bibitem [{\citenamefont {Walraven}\ \emph {et~al.}(2024)\citenamefont
  {Walraven}, \citenamefont {Tarbutt},\ and\ \citenamefont
  {Karman}}]{walraven:24b}%
  \BibitemOpen
  \bibfield  {author} {\bibinfo {author} {\bibfnamefont {E.~F.}\ \bibnamefont
  {Walraven}}, \bibinfo {author} {\bibfnamefont {M.~R.}\ \bibnamefont
  {Tarbutt}},\ and\ \bibinfo {author} {\bibfnamefont {T.}~\bibnamefont
  {Karman}},\ }\href {https://doi.org/10.1103/PhysRevLett.132.183401}
  {\bibfield  {journal} {\bibinfo  {journal} {Phys. Rev. Lett.}\ }\textbf
  {\bibinfo {volume} {132}},\ \bibinfo {pages} {183401} (\bibinfo {year}
  {2024})}\BibitemShut {NoStop}%
\bibitem [{\citenamefont {Kaufman}\ and\ \citenamefont
  {Ni}(2021)}]{kaufman:21}%
  \BibitemOpen
  \bibfield  {author} {\bibinfo {author} {\bibfnamefont {A.~M.}\ \bibnamefont
  {Kaufman}}\ and\ \bibinfo {author} {\bibfnamefont {K.-K.}\ \bibnamefont
  {Ni}},\ }\href {https://doi.org/10.1038/s41567-021-01357-2} {\bibfield
  {journal} {\bibinfo  {journal} {Nature Phys.}\ }\textbf {\bibinfo {volume}
  {17}},\ \bibinfo {pages} {1324} (\bibinfo {year} {2021})}\BibitemShut
  {NoStop}%
\bibitem [{\citenamefont {Walraven}\ \emph {et~al.}(2026)\citenamefont
  {Walraven}, \citenamefont {Feng}, \citenamefont {Rodewald}, \citenamefont
  {Tarbutt},\ and\ \citenamefont {Karman}}]{tweezer}%
  \BibitemOpen
  \bibfield  {author} {\bibinfo {author} {\bibfnamefont {E.~F.}\ \bibnamefont
  {Walraven}}, \bibinfo {author} {\bibfnamefont {K.}~\bibnamefont {Feng}},
  \bibinfo {author} {\bibfnamefont {J.}~\bibnamefont {Rodewald}}, \bibinfo
  {author} {\bibfnamefont {M.~R.}\ \bibnamefont {Tarbutt}},\ and\ \bibinfo
  {author} {\bibfnamefont {T.}~\bibnamefont {Karman}},\ }\href
  {https://arxiv.org/abs/2607.25783} {\bibinfo {title} {Deterministic loading
  of molecular arrays by microwave-assisted collisions}} (\bibinfo {year}
  {2026}),\ \Eprint {https://arxiv.org/abs/2607.25783} {arXiv:2607.25783
  [physics.atom-ph]} \BibitemShut {NoStop}%
\end{thebibliography}%
\end{document}